\documentclass{emulateapj}

\usepackage{amsmath}

\newcommand{\be}{\begin{equation} }
\newcommand{\ee}{\end{equation} }
\newcommand{\bea}{\begin{eqnarray} }
\newcommand{\eea}{\end{eqnarray} }

\slugcomment{Submitted to ApJ: 14 Apr 2009, Accepted: 26 Oct 2009}

\shortauthors{Maness {\it et~al.\/}}

\shorttitle{HD 61005 Optical Imaging}

\begin{document}

\title{Hubble Space Telescope Optical Imaging of the Eroding Debris Disk HD 61005}

\author{ 
H. L. Maness\altaffilmark{1},
P. Kalas\altaffilmark{1},
K. M. G. Peek\altaffilmark{1},
E. I. Chiang\altaffilmark{1},
K. Scherer\altaffilmark{2},
M. P. Fitzgerald\altaffilmark{3}, 
James R. Graham\altaffilmark{1},
D. C. Hines\altaffilmark{4},
G. Schneider\altaffilmark{5},
S. A. Metchev\altaffilmark{6}
} 

\altaffiltext{1}{Department of Astronomy, University of California at
Berkeley, Berkeley, CA 94720}
\altaffiltext{2}{Institut f\"ur Theoretische Physik IV, Ruhr Universit\"at
  Bochum, 44780 Bochum, Germany}
\altaffiltext{3}{Lawrence Livermore National Laboratory, Livermore, CA
94550}
\altaffiltext{4}{Space Science Institute, Boulder, CO 80301}
\altaffiltext{5}{Steward Observatory, University of Arizona, Tucson,
  AZ 85721}
\altaffiltext{6}{Department of Physics and Astronomy, State University
  of New York $-$ Stony Brook, Stony Brook, NY 11794}

\begin{abstract}

  We present Hubble Space Telescope optical coronagraphic polarization
  imaging observations of the dusty debris disk HD 61005.  The
  scattered light intensity image and polarization structure reveal a
  highly inclined disk with a clear asymmetric, swept back component,
  suggestive of significant interaction with the ambient interstellar
  medium.  The combination of our new data with the published 1.1
  $\mu$m discovery image shows that the grains are blue scattering
  with no strong color gradient as a function of radius, implying
  predominantly sub-micron sized grains.  We investigate possible
  explanations that could account for the observed swept back,
  asymmetric morphology.  Previous work has suggested that HD 61005
  may be interacting with a cold, unusually dense interstellar
  cloud. However, limits on the intervening interstellar gas column
  density from an optical spectrum of HD 61005 in the Na I D lines
  render this possibility unlikely. Instead, HD 61005 may be embedded
  in a more typical warm, low-density cloud that introduces secular
  perturbations to dust grain orbits.  This mechanism can
  significantly distort the ensemble disk structure within a typical
  cloud crossing time.  For a counterintuitive relative flow
  direction---parallel to the disk midplane---we find that the
  structures generated by these distortions can very roughly
  approximate the HD 61005 morphology.  Future observational studies
  constraining the direction of the relative interstellar medium flow
  will thus provide an important constraint for future modeling.
  Independent of the interpretation for HD 61005, we expect that
  interstellar gas drag likely plays a role in producing asymmetries
  observed in other debris disk systems, such as HD 15115 and $\delta$
  Velorum.

\end{abstract}


\section{Introduction}

Nearly two dozen dusty debris disks surrounding nearby stars have now
been spatially resolved at one or more wavelengths.  Many of these
systems show clear similarities.  For example, the radial architecture
of several debris disks can be understood in terms of a unified model
of steady-state dust production via collisions in a parent
planetesimal belt (e.g., Strubbe \& Chiang 2006).  However, while the
observed structure of many systems is ring-like
\citep{Kalas06,Wyatt08}, most disks show substructure such as clumps,
warps, offsets, and brightness asymmetries not explained in
traditional steady-state collisional grinding models.

These unexpected features have triggered a great deal of recent
theoretical work.  The effects of massive planetesimal collisions,
sandblasting by interstellar grains, close stellar flybys, dust avalanches, and
secular and resonant perturbations by exoplanets have all been invoked
to explain the observations (e.g., Moro-Martin et al. 2007, and references
therein).  However, as many of these theories produce similar
structures, further observational constraints are needed to better
understand the key forces affecting disk structure and the
circumstances in which they apply.

At a heliocentric distance of 34.5 pc \citep{Perryman97}, the debris
disk surrounding HD 61005 (SpT: G8 V; Gray et al. 2006), is a
promising target for advancing our understanding in this area.  The
significant infrared excess for this source ($L_{\rm IR}/L_{*} = 2
\times 10^{-3}$) was recently discovered as part of the Spitzer FEPS
survey \citep{Carpenter09}, indicating 60 K blackbody-emitting grains
$\gtrsim$ 16 AU from the star.  Follow-up Hubble Space Telescope (HST)
coronagraphic imaging observations with the Near Infrared Camera and
Multi-Object Spectrometer (NICMOS; HST/GO program 10527; D. Hines, PI)
resolved the source (Hines et al. 2007, hereafter H07), revealing an
unprecedented swept, asymmetric morphology, suggestive of significant
interaction with the interstellar medium (ISM).  H07 suggested that
this system could be a highly inclined debris disk, undergoing ram
pressure stripping by the ambient ISM.  However, this interpretation
requires an unusually high interstellar density for the low-density
Local Bubble in which HD 61005 resides.  Furthermore, the single
wavelength intensity image was insufficient to provide strong
constraints on the dominant size of the scattering grains and the
overall scattering geometry.

To further quantify the physical properties of grains seen in
scattered light and the overall geometry of the system, we obtained
optical coronagraphic polarimetry imaging observations of HD 61005
with the Advanced Camera for Surveys (ACS) onboard HST.  As
demonstrated by \citet{Graham07} for the case of AU Mic, polarization
observations in scattered light are invaluable for breaking
degeneracies between grain scattering properties and their spatial
distribution.  Furthermore, the ACS data represent a factor of two
improvement in angular resolution compared to the 1.1 $\mu$m discovery
observations.  In addition to these new imaging data, we also obtained
a high resolution optical spectrum to characterize ambient
interstellar gas surrounding this system.  In \S 2, we describe the
steps taken in observing and reducing these data.  In \S 3, we discuss
the results of these observations, their consequences for the system
scattering geometry, and the additional constraints they provide when
combined with the 1.1 $\mu$m NICMOS image.  In \S 4, we explore
whether interactions with ambient interstellar gas can plausibly
explain the observed swept, asymmetric morphology in this system.  We
discuss the implications for these potential explanations in \S 5 and
summarize our findings in \S 6.

\section{Observations and Data Reduction}

We obtained optical coronagraphic observations of HD 61005 using the
ACS high resolution camera (HRC) $1\farcs8$ diameter occulting spot on
2006 December 19 (HST/GO program 10847; D. Hines, PI).  In each of two
contiguous orbits, we imaged HD 61005 with the F606W filter in
combination with the POL0V, POL60V and POL120V polarizer filters (two
340 second exposures per filter combination).  Aside from the
telescope position angle, which is rotated 23.032 degrees between
orbits, the observational procedures for the two HD 61005 orbits were
identical.  Before the HD 61005 orbits, we observed the
point-spread function (PSF) reference star HD 82943 (SpT: F9 V; Gray
et al. 2006; B = 7.16, V = 6.56) using an observing sequence identical
to the two orbits allocated to HD 61005 (B = 8.93, V = 8.22).  We also
observed a second PSF star, HD 117176 (SpT: G5 V; Gray et al. 2001; B
= 5.69, V = 5.00), in an identical manner following the HD 61005
orbits.

For each filter combination, we combined the two 340 second, pipeline
processed (bias subtracted, flatfielded) frames by excluding the
maximum value at each pixel position, thereby minimizing the impact of
cosmic ray events.  After dividing by the cumulative integration time
of each frame, we performed sky subtraction by taking the median value
in a 10$\times$20 pixel box in the lower left corner of the chip,
which is the position farthest from the bright target star.  We
registered the images by selecting a fiducial HD 61005 image (the
POL0V image in the first orbit) and subtracting all other HD 61005
frames using small offsets (0.02 pixels) to minimize the residuals in
regions dominated by light from the stellar PSF.  The offsets that
minimize residual differences between frames were then applied to the
individual POL0V, POL60V, and POL120V images to align them to a common
reference frame relative to the star.  We carried out an identical
registration procedure for the two PSF reference stars.

We then subtracted the HD 61005 PSF in each of the three POL0V,
POL60V, and POL120V frames by the corresponding frames from each of
the two PSF reference stars.  Prior to subtraction, we scaled each
reference star to match the expected brightness of HD 61005, using
photometry obtained from the direct images.  The HD 117176
observations, made immediately following the HD 61005 orbits yielded a
better subtraction than the HD 82943 observations, made five weeks
prior to the HD 61005 orbits.  We therefore used the subtraction
obtained with HD 117176 for all subsequent analysis.

Following PSF subtraction, we corrected the resultant images for
geometric distortion yielding 25 mas $\times$ 25 mas pixels.  We then
constructed Stokes parameter images corrected for instrumental
polarization following \citet{Pavlovsky06}.  For the
ACS/HRC/F606W/POLV instrumental configuration and high fractional
linear polarization ($p = (Q^2 + U^2)^{1/2}/I \ge 0.2$), the residual
systematic error is 10\% of the computed polarization fraction.  For
less strongly polarized sources ($p < 0.2$), the systematic error in
the degree of linear polarization is approximately constant at
$\sigma_{p} = 0.01$.  In both cases, the systematic uncertainty in
position angle is $3^\circ$.

We next calculated polarization vectors from the derived Stokes
images.  As the polarization fraction is intrinsically positive and
biased upwards by noise, we employed the spatial binning algorithm of
\citet{Cappellari03} to bin the Stokes $I$, $Q$, and $U$ images to
approximately constant signal-to-noise prior to this
calculation. Within $\sim 1.4^{\prime\prime}$ of the star, the computed polarization
vectors become significantly disordered in magnitude and direction, as
a result of systematic PSF subtraction errors. We therefore only
consider polarization vectors outside this radius in our analysis.
The surface brightness at $0.9-1.4^{\prime\prime}$ is similarly compromised and
should be treated with caution.

Finally, we converted from instrumental to physical brightness units
using the synthetic photometry package, Synphot.  As input, Synphot
requires the instrument configuration (camera, coronagraph, wideband
and polarizing filters) and the source spectrum across the band.
Since the latter is unknown, we performed the calculations three times
assuming: (1) a flat spectrum, (2) a $T_{\rm eff} = 5500$ K Kurucz
synthetic spectrum approximating the G-dwarf stellar spectrum, and (3)
a spectral slope across each band that is the same as that calculated
between the NICMOS and ACS bands using method (2).  All methods
yielded conversion factors within 1\% of each other, suggesting the
assumed source spectrum factors negligibly into the total color
uncertainties.

In addition to the ACS observations, we analyze two additional
data sets: (1) The NICMOS F110W image; a full description of
the NICMOS data acquisition and reduction is given in H07. (2) High
resolution (R $\approx$ 60,000) echelle spectra for HD 61005 and two
comparison stars of similar spectral type (HD 33822: $T_{\rm eff} =
5850$ K, HD 13836: $T_{\rm eff} = 5580$ K; Masana et al. 2006).  The
spectra were obtained on the Keck I telescope with the HIRES
spectrometer on 30 Dec 2004, 16 Jan 2006, and 09 Nov 2008.  The
wavelength range was 3700-6200 \AA, though our analysis concerns only
the Na I D lines at 5889.951 {\AA} and 5895.924 \AA.  We used a standard
procedure to perform flat-fielding, sky subtraction, order extraction,
and wavelength calibration of the raw echelle images \citep{Butler96,
  Vogt94}.

\section{Results}

\subsection{ACS Scattered Light and Polarization}

\subsubsection{Two-dimensional morphology and polarization structure}

\begin{figure}
\epsscale{0.3}
\includegraphics[width=0.37\textwidth,angle=90]{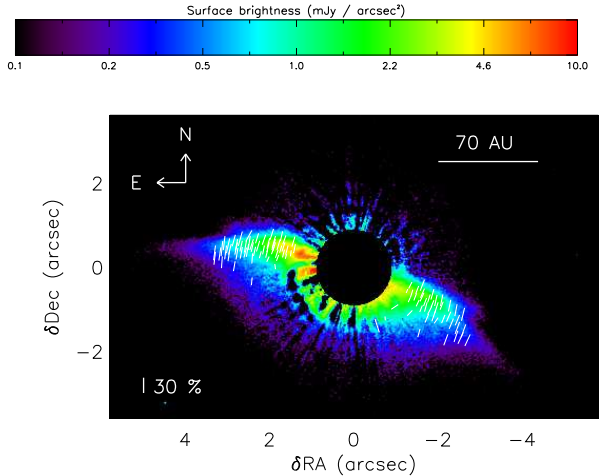} 
\caption{Logarithmically-scaled image of HD 61005 in Stokes $I$ at
  $\ge 0.1$ mJy arcsec$^{-2}$ with polarization vectors and a 1.8$^{\prime\prime}$
  coronagraphic mask overplotted.  The color bar units were calculated
  using the synthetic photometry package, Synphot.  The plotted
  polarization vectors were computed from Stokes images binned to
  approximately constant signal-to-noise.}
\label{acs_image_log}
\end{figure}

\begin{figure}
\epsscale{0.3}
\includegraphics[width=0.37\textwidth,angle=90]{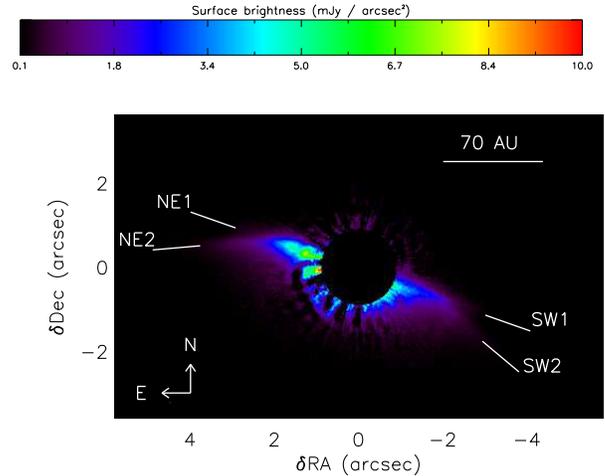} 
\caption{Same as Figure \ref{acs_image_log}, only with a linear
  display scale and without polarization vectors.  Overplotted are
  labels for the surface brightness components displayed in Figure
  \ref{sbp}.}
\label{acs_image_lin}
\end{figure}

Figures \ref{acs_image_log} and \ref{acs_image_lin} display the F606W
total intensity image of HD 61005 on a logarithmic and linear scale,
respectively.  The figures show two distinct morphological components.
The first component, denoted by NE1 and SW1 in Figure
\ref{acs_image_lin}, resembles a near edge-on disk.  The putative
midplane for this component is observed to extend out to $\sim 3^{\prime\prime}$
from the star, where the signal-to-noise per pixel falls below unity.
The second morphological component is detected below the nominal disk
midplane.  This component exhibits an asymmetric, ``swept back''
morphology, suggestive of significant interaction with the
interstellar medium.  This unusual structure was first noted by H07 in
their NICMOS F110W discovery image.  Both the disk-like and swept back
components additionally exhibit a striking asymmetry between the
northeast and southwest sides of the source.  At a given projected
radius, the northeast side of the source is approximately twice as
bright as the southwest side.  This brightness asymmetry is also seen
in the NICMOS F110W image (e.g., H07, Figure 4).

\begin{figure}
\epsscale{0.3}
\includegraphics[width=0.37\textwidth,angle=90]{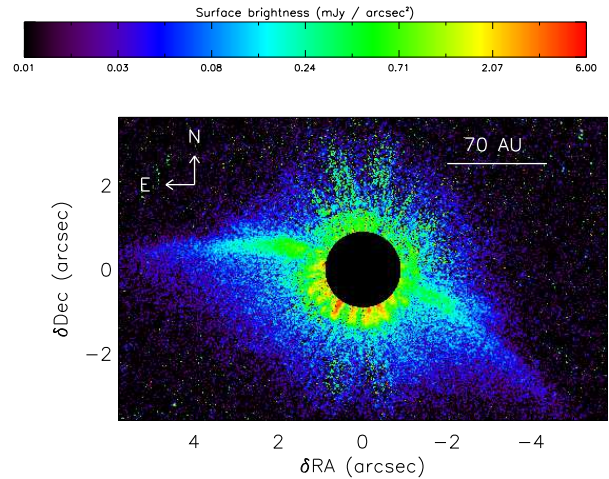} 
\caption{Logarithmically-scaled polarized flux ($\sqrt{Q^2+U^2}$)
  image of HD 61005.}
\label{moth_ip}
\end{figure}

Figure \ref{moth_ip} displays the polarized flux of image of HD 61005,
while Figure \ref{acs_image_log} overplots polarization vectors on the
Stokes $I$ image, spatially binned according to the procedure outlined
in \S 2.  The polarization vectors in Figure \ref{acs_image_log} show
that along the plane of the disk, the fractional polarization
increases with radial distance from the star from $\sim 10$\% to $\sim
35$\%.  The orientation of the electric field within $\sim 2.1^{\prime\prime}$ also
appears perpendicular to the nominal disk midplane.  In the swept back
area of the source, the polarization vectors rotate to become
approximately perpendicular to the outer edge of this component.  The
fractional polarization in this region is similar to that seen in the
outer part of the disk component.

\subsubsection{Disk component parameters}

The ACS polarization results add further evidence to the suggestion by
H07 that HD 61005 is a near edge-on debris disk.  In particular, the
HD 61005 polarization structure is very similar to that seen in the
spectrally blue edge-on debris disk around the M dwarf, AU Mic,
observed using the same instrumental configuration \citep{Graham07}.
The midplanes of both HD 61005 and AU Mic exhibit high fractional
polarization ($p_{\rm max} \sim 0.4$) which increases with projected
radius.  Similarly, both disks exhibit an electric field orientation
perpendicular to the disk midplane at all projected radii.  These
features are expected for small-particle ($x \lesssim 1$) scattering
in an edge-on disk \citep{Kruegel03}.

\begin{figure}
\epsscale{0.3}
\includegraphics[width=0.47\textwidth,angle=0]{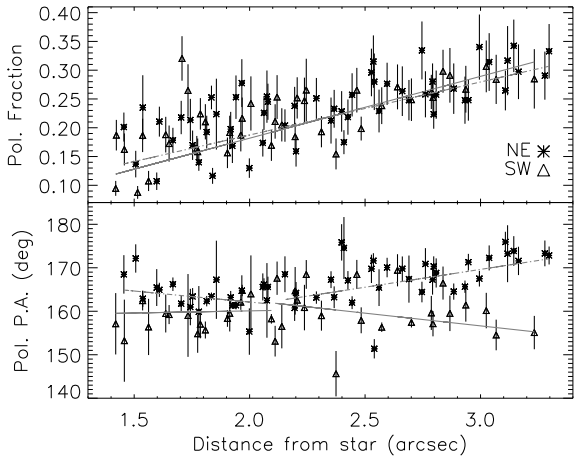} 
\caption{One dimensional version of the binned polarization vectors
  displayed in Figure \ref{acs_image_log}.  In the top panel, the data
  show a clear increase in fractional polarization with distance from
  the star. Linear fits to the data are overplotted in grey with a
  solid line for the southwest side and a dot-dashed line for the
  northeast side.  In the bottom panel, the rotation of the electric
  field orientation in the swept back outer disk is seen in the
  deviation from the nominal midplane polarization position angle
  outside $\sim2.1^{\prime\prime}$.  Linear fits for data in and
  outside this radial distance are overplotted. \vspace*{3mm}}
\label{1d_pol}
\end{figure}

For HD 61005, these effects are quantified in Figure \ref{1d_pol},
which shows the binned polarization vector position angles and
magnitudes from Figure \ref{acs_image_log}.  Linear fits to the
one-dimensional polarization position angles (PA) versus projected
distance within $2.1^{\prime\prime}$ imply flat slopes, as expected
for a highly-inclined disk geometry ($-5.1 \pm 2.5$ and $1.0 \pm 3.3$
for the northeast and southwest sides, respectively).  Averaging all
polarization position angles within $2.1^{\prime\prime}$, the implied
disk position angle is $71.7 \pm 0.7$ deg, where the listed
uncertainty is purely statistical and does not include the additional
3 deg calibration uncertainty (\S 2).  The steady increase in
polarization fraction with projected radius is also seen in Figure
\ref{1d_pol}.  Linear fits to the binned polarization fraction over
the full extent of the source imply slopes of $0.09 \pm 0.01$ and
$0.11 \pm 0.01$ for the northeast and southwest disk sides,
respectively.

As an independent check on our disk interpretation, we fit elliptical
isophotes to the total intensity image from Figures
\ref{acs_image_log} and \ref{acs_image_lin}, assuming a circularly
symmetric disk viewed in projection.  We independently fit eight
isophotes outside $1.4^{\prime\prime}$ for $I>1.3$ mJy arcsec$^{-2}$.  The implied
disk position angle from these fits is $70.7 \pm 0.5$ deg, in good
agreement with the position angle inferred from the electric field
orientation within $2.1^{\prime\prime}$.  The implied inclination to the line of sight
from these fits is $i = 80.3 \pm 0.6$ deg.

\subsubsection{Swept component parameters}

The disk structure described above contrasts with previously imaged
presumed interstellar dust phenomena, such as the infrared bow
structure surrounding the A star $\delta$ Velorum \citep{Gaspar08} and
the filamentary cirrus surrounding some Vega-like stars with
significant infrared excess \citep{Kalas02}.  Nevertheless, Figures
\ref{acs_image_log} and \ref{acs_image_lin} clearly reveal a second
asymmetric component of the source not typical of nearby debris disks
and suggestive of interaction with the interstellar medium.

The polarization signature of this component is evident in the
systematic rotation of the polarization vectors outside $\sim 2.1^{\prime\prime}$.
Linear fits to the one-dimensional polarization position angles versus
projected distance (Figure \ref{1d_pol}) give non-zero slopes of $8.2
\pm 1.0$ for the northeast lobe $-5.8 \pm 1.9$ for the southwest lobe.
Estimates for the position angles of the outer edge of this component
(cuts NE2 and SW2 in Figure \ref{acs_image_lin}) are obtained from the
outermost polarization vector position angles.  The implied position
angle for component NE2 is $82.8 \pm 2.1$ deg; the result for SW2 is
$65.1 \pm 3.8$ deg.

\subsubsection{One-dimensional surface brightness profiles}
\label{sbp_section}

\begin{figure*}
\epsscale{0.3}
\begin{center}
\includegraphics[width=0.7\textwidth,angle=0]{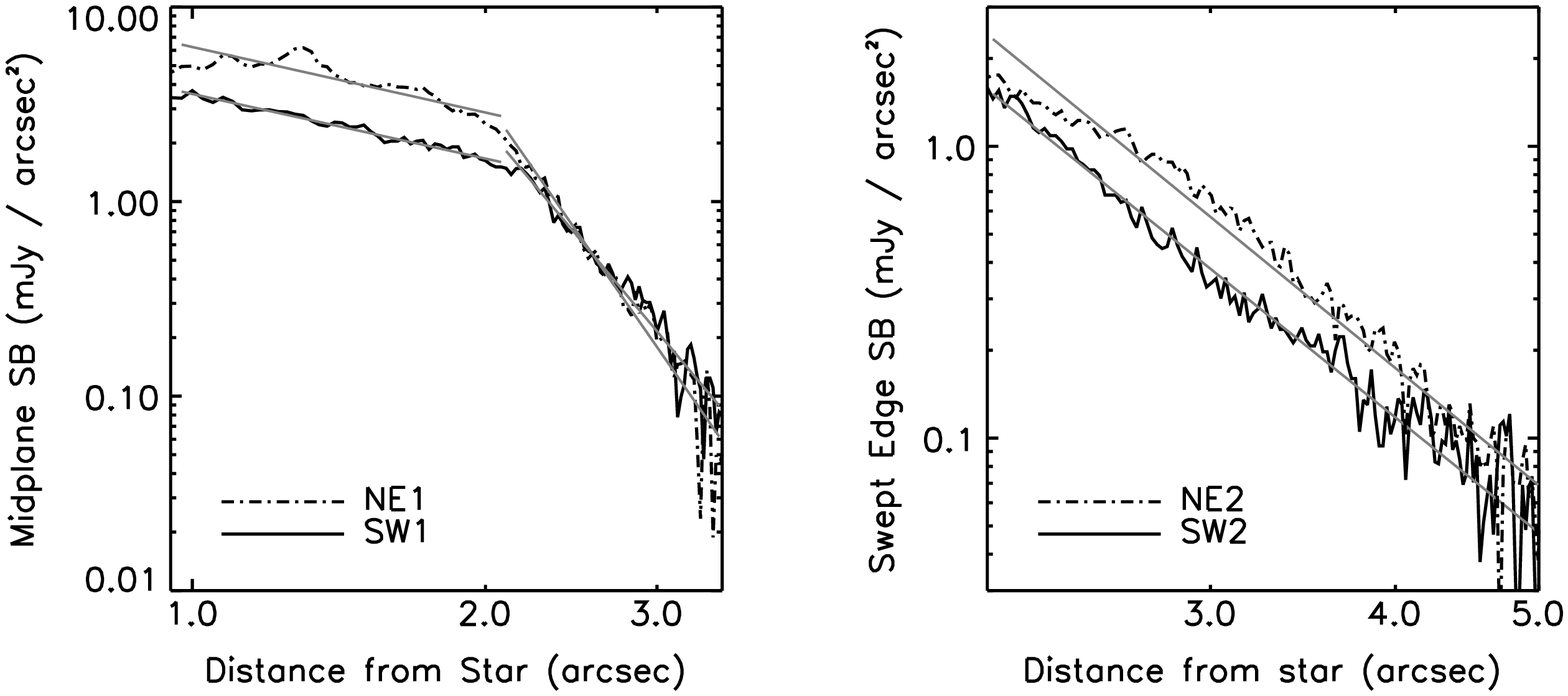} 
\end{center}
\caption{Surface brightness profiles through the disk midplane (left)
  and the outer edge of the swept back component (right).  The cuts
  are one pixel wide, and the components listed in the legends are
  labeled in Figure \ref{acs_image_lin}.  The directions for the cuts
  are taken from the position angles derived in \S 3.1.2 and
  3.1.3. The power law fits described in the text are overplotted in
  grey.  The data show a pronounced brightness asymmetry between the
  northeast and southwest disk lobes.}
\label{sbp}
\end{figure*}

One-dimensional surface brightness profiles through the disk midplane
(components NE1 and SW1 in Figure \ref{acs_image_lin}; PA = $70.7$
deg) and along the swept component outer edge (components NE2 and SW2
in Figure \ref{acs_image_lin}; PA = $82.8$ deg, $65.1$ deg) are shown
in Figure \ref{sbp}.  The midplane surface brightness follows a broken
power law.  The fitted power law indices between 0.9$^{\prime\prime}$ and 2.1$^{\prime\prime}$ are
$-1.1 \pm 0.1$ for both midplane disk lobes.  Between 2.1$^{\prime\prime}$ and
3.5$^{\prime\prime}$, power law fits give indices of $-7.2 \pm 0.3$ for component
NE1 and $-6.0 \pm 0.2$ for component SW1.  Breaks in the
scattered-light surface brightness of debris disks are often taken as
the location of the parent body population for the scattering grains
(e.g., Strubbe \& Chiang 2006).  However, as the midplane break
approximately coincides with the position at which the outer edge of
the swept component intersects the disk, this standard interpretation
may not hold for this case.

The surface brightness profile for the outer edge of the swept
component is well represented by a single power-law (Figure
\ref{sbp}).  The fitted power law index is $-4.1 \pm 0.1$ for both
NE2 and SW2.  The listed errors for the power law fits in this section
are formal fitting errors and should be treated as lower limits to the
true uncertainties.

\subsection{ACS+NICMOS}

\subsubsection{Disk Scattered Light Colors}

We computed the color of the disk by rebinning the ACS Stokes $I$
image to the same pixel resolution as the NICMOS image
(0.0759$^{\prime\prime}$).  We then convolved the binned ACS image
with a coronagraphically unocculted field star (the approximate NICMOS
PSF) and performed the corresponding operation on the NICMOS image.
Finally, we divided each image by the stellar flux density of HD 61005
at the appropriate band-center effective wavelength before dividing
the NICMOS image by the ACS image.

\begin{figure*}
\begin{center}
\includegraphics[width=0.8\textwidth,angle=0]{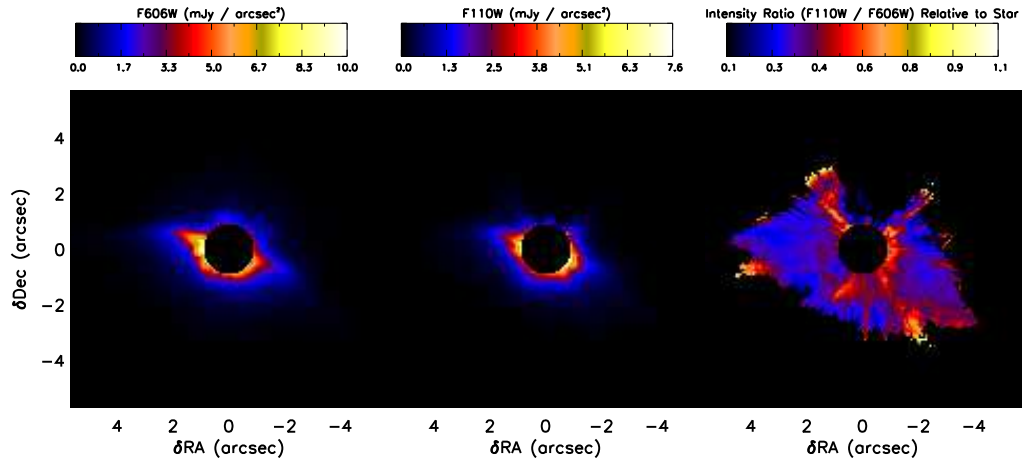} 
\end{center}
\caption{ACS and NICMOS images processed to compute the grain color.
  The left panel displays the ACS image binned to the NICMOS pixel
  resolution and convolved with the NICMOS off-spot PSF.  The middle
  panel displays the NICMOS image convolved with the ACS PSF.  The
  right panel shows a masked ratio image of the left and middle images
  divided by the stellar flux density ratio; values less than unity
  represent grains that preferentially scatter blue light, whereas
  values greater than unity represent grains that preferentially
  scatter red light.  In the ratio image, we have applied a mask to
  all pixels with values less than 2.5 times the background level in
  either original image. The ratio image indicates that the disk
  appears predominantly blue with no appreciable color gradient.}
\label{color}
\end{figure*}

Figure \ref{color} illustrates that the HD 61005 debris disk appears
predominantly blue with no significant systematic color gradient.  The
mean intensity ratio inferred from Figure \ref{color} is $0.32 \pm
0.10$ (corresponding to a color index of [F606W]$-$[F110W] = $-1.2 \pm
0.3$), where the error is dominated by uncertainties in the PSF
subtraction.  Although comparing images of different resolution can
result in systematic color errors (e.g., Golimowski et al. 2006), our
convolution steps appear to have a small effect; the mean color
neglecting convolution is within $1\sigma$ of that found including
convolution.

The blue color inferred from Figure \ref{color} (ratio $< 1$) is rare,
as the handful of debris disks with color measurements to date show
mainly red colors (Meyer et al. 2007, and references therein).  There
are several notable exceptions, however.  The HD 32297 and HD 15115
debris disks, for example, have been suggested to show blue optical to
near-infrared scattered light colors \citep{Kalas05,Kalas07}, though
the result for HD 32297 is currently under debate \citep{Debes09}.
Interestingly, both disks have morphological features consistent with
ISM interaction.  HD 32297 shows a bowed disk structure, similar to HD
61005, though on a much larger scale ($\sim$ 1000 AU; Kalas et
al. 2005). HD 15115 is highly asymmetric, perhaps as the result of ISM
erosion \citep{Debes09}.

The M dwarf debris disk, AU Mic, which has a similar polarization
structure to HD 61005 (\S 3.1.2), also shows blue optical to
near-infrared colors, with a color gradient towards bluer colors at
larger radial distances, indicating changes in the grain size
distribution \citep{Strubbe06,Fitzgerald07}.  The global
[F606W]$-$[F110W] color of the HD 61005 disk is comparable to the
[F606W]$-J$ AU Mic disk color at projected radii within $\sim 40$ AU,
the approximate location of the parent body ring in this system
\citep{Fitzgerald07}.

\begin{figure}
\begin{center}
\includegraphics[width=0.43\textwidth,angle=0]{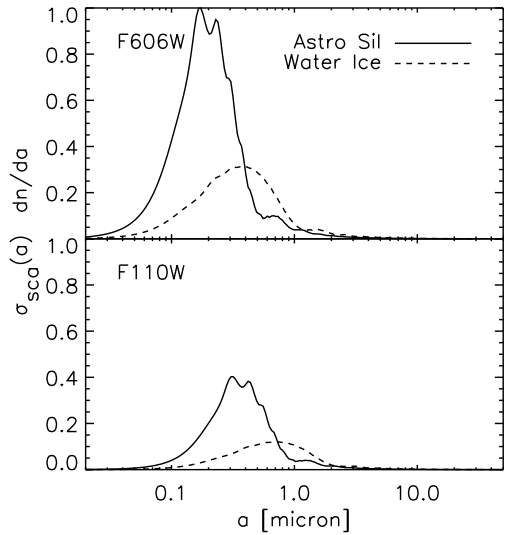} 
\end{center}
\caption{Product of the scattering cross section and a Dohnanyi size
  spectrum as a function of logarithmic grains radius for spherical
  particles.  At a given wavelength, this product largely determines
  the surface brightness at a particular location in the disk.
  Assuming a steep size spectrum (e.g., Dohnanyi), a large peak is
  observed near $x \sim 1$.  Thus a given scattered light image traces
  a relatively narrow window of the grain size population.  The HD
  61005 disk is globally brighter at optical wavelengths than
  near-infrared wavelengths, likely reflecting the larger number of
  grains at increasingly small sizes.  The above results are
  monochromatic: $\lambda_{\rm F606W} = 0.6\ \mu{\rm m}$ and
  $\lambda_{\rm F110W} = 1.1\ \mu$m.}
\label{cross_section}
\end{figure}

Like AU Mic, the blue color of the HD 61005 disk is likely due to the
disk grain size distribution.  Scattered light images afford a
relatively narrow window on the grain size population because the
brightness at a given location in the disk is largely determined by
the product of the grain size distribution and the grain scattering
cross section.  In the presence of a steep size spectrum
characteristic of a collisional cascade, this product is strongly
peaked near $x \equiv \frac{2 \pi a}{\lambda} \sim 1$.  For example,
for the Dohnanyi spectrum with $dn/da \propto a^{-7/2}$
\citep{Dohnanyi69}, this peak occurs at $x \simeq 2-6$, depending on
the optical properties of the grain material.  In Figure
\ref{cross_section}, a Mie calculation shows the peak for water ice
and astronomical silicates at $0.6\, \mu$m and $1.1\, \mu$m as a
function of grain size.  The plots illustrate that the HD 61005 NICMOS
images trace grains with radii of order $0.2-2\, \mu$m, while the ACS
images trace systematically smaller grains with radii of order
$0.1-1\, \mu$m.  Thus the observation that HD 61005 is globally
brighter at optical wavelengths than near-infrared wavelengths
suggests the disk contains a larger number of grains at increasingly
small sizes, consistent with our expectation of a steep size spectrum.

The inference that the F606W ACS images trace predominantly sub-micron
sized grains is also consistent with the imaging polarimetry results
(\S 3.1).  The large polarization fraction and electric field
orientation perpendicular to the edge-on midplane are in qualitative
agreement with the expected signature of scattering by small spherical
particles with $x \lesssim 1$ \citep{Kruegel03}.  For
larger spherical grains, the electric field orientation
can rotate by $90^\circ$ at certain scattering angles, resulting in an
orientation parallel to the edge-on midplane.  Furthermore, any line
of sight comprising emission from a range of scattering angles will
tend to show weak linear polarization \citep{Graham07}.  Neither of
these features is consistent with the HD 61005 polarization
results.

\begin{figure}
\includegraphics[width=0.47\textwidth,angle=0]{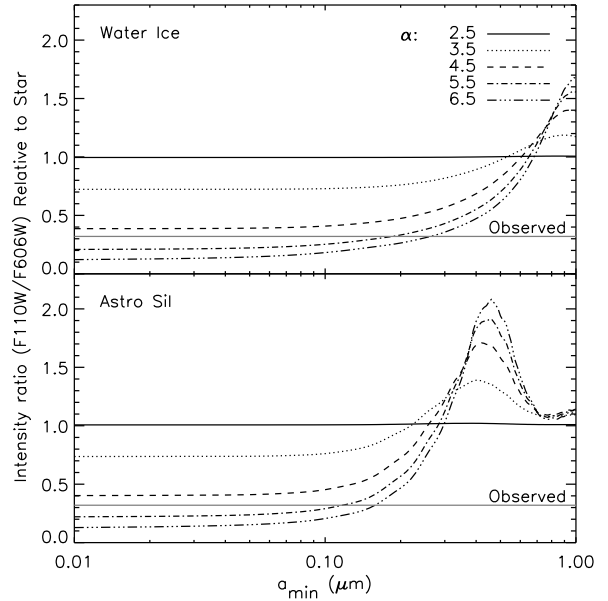} 
\caption{Implied disk colors for ice and silicate grains for grain
  size distributions of the form: $dn/da \propto a^{-\alpha}$, where
  $\ a_{\rm min} < a < 1\ {\rm mm}$.  The data place an upper limit on
  the minimum grain size ($a_{\rm min} \lesssim 0.3\ \mu$m) and
  suggest a global size distribution steeper than the canonical
  Dohnanyi size spectrum ($\alpha = 3.5$).  }
\label{color_calculation}
\end{figure}

By integrating the curves in Figure \ref{cross_section} and comparing
the results to the measured color, we can constrain the size
distribution.  The results of this procedure are shown in Figure
\ref{color_calculation}, which shows the implied color for ice and
silicate grains for grain size distributions of the form: $dn/da
\propto a^{-\alpha}$, where $\ a_{\rm min} < a < 1\ {\rm mm}$.  A
range of minimum grain sizes are considered, as radiation pressure
could potentially remove a fraction of small grains from the system
(Appendix A).  The data place an upper limit on the minimum grain
size: $a_{\rm min} \lesssim 0.3\ \mu$m.  Figure
\ref{color_calculation} also suggests the global size distribution is
steeper than the Dohnanyi spectrum: $\alpha \sim 4.5 - 5.5$.  This
slope is consistent with results from collisional equilibrium modeling
of other debris disk systems (e.g., Strubbe \& Chiang 2006). However,
the applicability of these models to the unique HD 61005 system
is presently uncertain.

\subsubsection{Astrometric test for low-mass companions}

The ACS and NICMOS images contain four field stars in common to both
data sets, roughly offset from HD 61005 in RA and Dec by
[-2.4$^{\prime\prime}$, 9.4$^{\prime\prime}$ ],
[-8.6$^{\prime\prime}$, 0.0$^{\prime\prime}$ ],
[-12.8$^{\prime\prime}$, 3.5$^{\prime\prime}$], and
[-14.0$^{\prime\prime}$, 2.4$^{\prime\prime}$].  As the ACS and
first-epoch NICMOS images were obtained 1.1 years apart, we can
measure the field star proper motions to investigate whether any of
these stars are likely companions to HD 61005.  The annular proper
motion of HD 61005 is $\mu_\alpha = -56.09$ mas yr$^{-1}$ in right
ascension and $\mu_\delta = 74.53$ mas yr$^{-1}$ in declination
\citep{Perryman97}. Thus comoving companions are expected to show a
102.61 mas displacement between the two epochs of observation.

To derive the NICMOS stellar positions, we first calculated stellar
centroid positions on the image frames uncorrected for geometric
distortion using the apphot $center$ task, as described in
\citet{Cox97}.  We then applied a distortion correction to the raw
positions using the correction coefficients appropriate for Cycle 15
data, given in the NICMOS data handbook (version 7.0).  The reported
rms astrometric uncertainty from applying previously derived
distortion corrections to early commissioning data is 13.6 mas
\citep{Cox97}.  Data obtained more recently yield similar results
(e.g., Schneider et al. 2006).

The geometric distortion for the ACS data requires a low spatial
frequency correction for the optical telescope assembly (OTA) and ACS
optics, and two high frequency corrections for the given wide band
filter and polarizing filter \citep{Anderson04,Kozhurina04}.
Following \citet{Kozhurina04}, we first used the effective PSF library
and fitting technique of \citet{Anderson00} to derive raw positions
for the field stars from the flat-fielded images ($*\_flt.fits$)
observed through the POL0V filter.  We next applied the solution
derived by \citet{Anderson04} to obtain stellar positions corrected
for the low frequency OTA distortion and the high frequency F606W
filter distortion.  Finally, we applied a further correction for the
distortion introduced by the POL0V filter, using the solution derived
by \citet{Kozhurina04}.  The reported rms precision derived from
applying this method to commissioning data is 1 mas
\citep{Kozhurina04}.

To test for possible companionship, we performed relative astrometry
by adopting one star's position as a fixed reference point and
calculating the relative proper motion for the remaining three stars.
We repeated this procedure three times, using each field star as the
reference.  No systematic motion for any of the field stars is
observed; all relative proper motions are less than twice the
approximate expected rms positional accuracy ($\sigma$ = 13.6 mas).
We note that although the distortion solutions employed only strictly
apply to noncoronagraphic, direct imaging data, the small measured
astrometric offsets suggest the additional field distortions imposed
by the coronagraphic optics are negligible in this case.  Given that
the annual proper motion of HD 61005 is significantly greater than the
measured astrometric motions of all field stars, we conclude that
unless all four sources are companions, the four field stars are
background objects.

\section{Interpretation}

The results presented in H07 and the previous section strongly suggest
that HD 61005 is a near edge-on debris disk, undergoing significant
erosion by the ambient interstellar medium.  In this section, we
explore whether interaction between the disk and local interstellar
gas can plausibly explain the observed swept, asymmetric morphology.

\subsection{Interaction with a Cold, Dense Cloud}
\label{cold}

\subsubsection{Ram pressure stripping of bound grains}
\label{strip_bound}

In their discovery paper, H07 suggested that interaction with a cold
($T \sim 20$ K), dense ($n \sim 100$ cm$^{-3}$) cloud could
potentially explain the HD 61005 morphology.  In such a cloud, ram
pressure on disk grains from interstellar gas could unbind grains from
the system, analogous to the process that strips gas from cluster
galaxies \citep{Gunn72,vanGorkom04}.  

For ram pressure stripping to operate, the drag force on a grain must
be comparable to or greater than the gravitational force binding the
grain to the star. For a grain of radius $a_{\rm grain}$ and density
$\rho_{\rm grain}$ orbiting at distance $r$ from a star of mass
$M_{\rm star}$\footnote{Throughout this paper, we adopt $M_{\rm star}
  = 0.95M_{\odot}$, based on the pre-main-sequence evolutionary
  tracks of \citet{Dantona97} and \citet{Baraffe98} and the FEPS age
  estimate reported by H07 (E. Mamejek, private communication).}, the
interstellar cloud density $n$ and relative cloud-disk velocity $v$
must obey:
\begin{equation}
\begin{split}
 \left( \frac{n}{200 \,{\rm cm}^{-3}} \right) \left( \frac{v}{30 \,{\rm km} \,{\rm s}^{-1}} \right)^2 \gtrsim 
 \left( \frac{M_{\rm star}}{0.95 \, M_\sun} \right)
 \left( \frac{a_{\rm grain}}{0.1 \, \mu {\rm m}} \right) \\
 \left( \frac{\rho_{\rm grain}}{2 \, {\rm g \, cm}^{-3}} \right)
 \left( \frac{70 \, \rm{AU}}{r} \right)^2.
\label{eq-ram-press}
\end{split}
\end{equation}
This required density is characteristic of cold, dense gas, the
existence of which is constrained in \S 4.1.3.  In such a high-density
cloud, H07 argued that Bondi-Hoyle-Lyttleton (BHL) accretion could
also potentially play a role, leading to an accumulation of
interstellar grains that contribute non-negligibly to the observed
infrared excess emission and scattered light morphology.

\subsubsection{Ram pressure deflection of unbound grains}
\label{deflect_unbound}

Recently, \citet{Debes09} suggested that ram pressure deflection of
unbound grains could plausibly shape several previously resolved
debris disks, including HD 61005.  Their model does not specify the
origin of the unbound population.  However, such a substantial
population of unbound grains are unlikely to be produced in a steady
state situation, as collisional equilibrium models predict that the
scattered light surface brightness due to bound grains dominates over
that from grains unbound by radiation pressure \citep{Krivov06,
  Strubbe06}.  Furthermore, for the case of HD 61005, radiation
pressure from the low luminosity star may be insufficient to unbind
grains of any size (Appendix A).

Let us nonetheless suppose that a substantial unbound grain population
exists. The ambient ISM density needed to deflect such grains by the
distances implied by our observations is similar to that required to
strip bound grains (\S \ref{strip_bound}). Both scenarios require
densities characteristic of cold, dense clouds.  For $\beta = 1$
grains launched from parent bodies on circular orbits at 70 AU, the
required interstellar cloud density and relative velocity
obey\footnote{Note that equations \ref{eq-ram-press} and
  \ref{eq-ram-press-ub} assume the cross section grains present to
  interstellar gas equals the geometric cross section; calculations
  performed using ballistic cluster-cluster aggregates and ballistic
  particle-cluster aggregates suggest this assumption is good to
  within an order of magnitude \citep{Minato06}.}:
\begin{equation}
\begin{split}
\left( \frac{n}{100 \,{\rm cm}^{-3}} \right) \left( \frac{v}{20 \,{\rm km} \,{\rm s}^{-1}} \right)^2 \gtrsim 
 \left( \frac{M_{\rm star}}{0.95 \, M_\sun} \right)
 \left( \frac{a_{\rm grain}}{0.1 \, \mu {\rm m}} \right) \\
 \left( \frac{\rho_{\rm grain}}{2 \, {\rm g \, cm}^{-3}} \right) 
 \left( \frac{70\, \rm{AU}}{r} \right) \\
 \left( \frac{y / x}{0.2} \right) 
 \left( \frac{110\, {\rm AU}}{x} \right).
\label{eq-ram-press-ub}
\end{split}
\end{equation}
Here $x$ and $y$, respectively, are the relative distances traveled by
the grain parallel and perpendicular to the disk midplane at a given
time after the grain is born.  We adopt $x \sim 110$ AU and $y \sim
20$ AU, the approximate positions of the outermost binned polarization
vectors in Figure \ref{acs_image_log}.

\subsubsection{Limits on cold, dense interstellar gas}
\label{cold_limit}

The explanations of \S \ref{strip_bound} and \S \ref{deflect_unbound}
are hampered by two factors.  First, in both circumstances, escaping
grains leave the system on orbital timescales ($10^2-10^3$ yr).  These
timescales are shorter than the timescale over which the disk is
expected to be collisionally replenished with sub-micron sized grains
($\gtrsim 10^4$ yr; see Appendix B).  Thus both scenarios require that
we are observing HD 61005 during a short-lived period in its history.

\begin{figure*}
\begin{center}
\includegraphics[width=0.7\textwidth,angle=0]{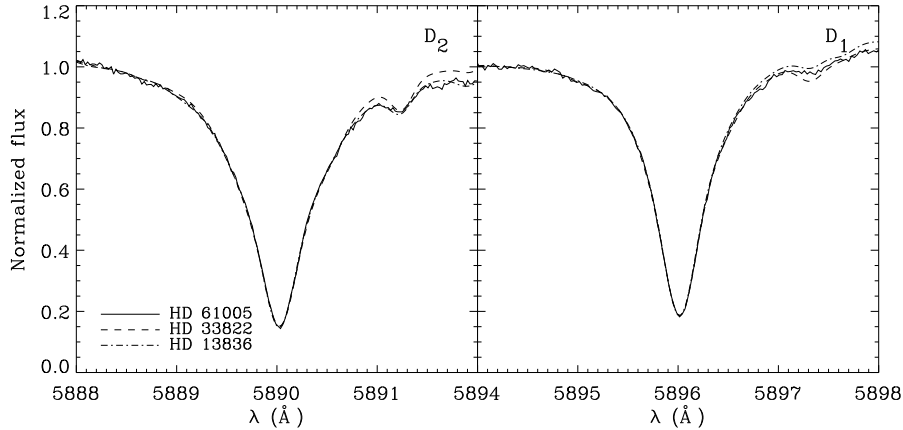} 
\end{center}
\caption{Optical spectrum of HD 61005 and two comparison late-type G
  star spectra in the Na I D lines.  The Na I lines in HD 61005 are
  very similar to the comparison star lines.  No narrow absorption
  component expected for an intervening dense cold interstellar cloud
  is observed, placing limits on the potential effectiveness of ram
  pressure stripping of bound grains and deflection of unbound grains
  in the HD 61005 system.}
\label{naI_spec}
\end{figure*}

Second, cold neutral medium (CNM) clouds within the Local Bubble are
extremely rare, occupying a volumetric filling factor of $\ll 1 \%$
\citep{Welsh94}.  To date, only one cloud with properties typical of
CNM clouds ($n \sim 50$ cm$^{-3}$, $ T\sim$ 20 K, $V =$ 4.9 $\times$
1.4 $\times$ 0.07 pc$^3$) is known within the Local Bubble ($d < 45$
pc, towards 3C 225; Heiles et al. 2003; Meyer et al. 2006; Meyer
2007).  Furthermore, in contrast to optical stellar spectra towards
this known cloud, our optical spectrum of HD 61005 does not show
evidence of an analogous CNM cloud. This finding is illustrated in
Figure \ref{naI_spec}, which shows spectra of HD 61005 and two
comparison late-type G stars (HD 33822, HD 13836) in the Na I D lines
(D$_2$: 5889.95 \AA, D$_1$: 5895.92 \AA).  HD 61005 is a relatively
young star with a detected rotational velocity (V$\sin i =$ 9 km
s$^{-1}$; Holmberg et al. 2008).  Thus for direct comparison purposes,
we have convolved the spectra of both HD 33822 and HD 13836 with a
Gaussian of this width.  In Figure \ref{naI_spec}, we have also
effectively removed the stellar Doppler shift of the comparison star
spectra by cross-correlating each spectrum with the HD 61005 spectrum
and shifting it by the appropriate amount.

Figure \ref{naI_spec} shows that the HD 61005 spectrum is very similar
to the late-G comparison star spectra in the Na I lines.  No narrow
absorption component, characteristic of the known CNM Local Bubble
cloud towards 3C 225, is observed (see Figure 1 in Meyer 2007).  The
corresponding 2$\sigma$ upper limit to the Na I gas column is $\log
(N_{\rm Na I}/ {\rm cm}^{-2}) \lesssim 10.6$, based on the continuum
signal-to-noise ratio (S/N $\sim$ 100), the instrumental resolution (R
$\sim$ 60,000), a linewidth characteristic of the nearby cold cloud
towards 3C 225 ($b = 0.54$ km s$^{-1}$), and assuming the linear
regime of the curve of growth.  This upper limit translates into a
total hydrogen column of $\log (N_{\rm HI+H_2}/ {\rm cm}^{-2})
\lesssim 19.0$, employing the sodium-hydrogen conversion relation
derived by Ferlet et al. (1985).  The scatter in this relation is such
that a third of measured hydrogen columns deviate from their
predicted columns by factors of a few \citep{Wakker00}.  Taking this
into account, our predicted hydrogen upper limit is still well below
the columns expected for CNM clouds ($10^{20} - 10^{21}$ cm$^{-2}$;
McKee \& Ostriker 1977) and also below the mean column density of Na I
detected towards 3C 225: $\langle \log (N_{\rm Na I} / {\rm cm}^{-2})
\rangle = 11.7$ \citep{Meyer06}.

There are a few examples of tiny ($\lesssim$ 1000 AU) low column
density ($\log (N_{\rm HI+H_2}/ {\rm cm}^{-2}) \lesssim 19.0$) cold
clouds in the literature (e.g., Heiles 1997, Stanimirovic \& Heiles
2005).  These clouds constitute $\lesssim 2-4$\% of the total neutral
hydrogen column along the lines of sight in which they are detected.
The precise frequency of these clouds is not well constrained by
large-scale surveys, as their low columns are similar to survey
detection thresholds ($\sim 10^{18}$ cm$^{-2}$; Heiles \& Troland
2005).  However, an upper limit to their frequency can be obtained by
noting that $\gtrsim$ 5 such clouds along a typical line of sight
would lead to their systematic detection in large-scale surveys
(C. Heiles, private communication).  Thus within the Local Bubble (d
$\lesssim 100$ pc), these tiny, dense clouds occupy $\lesssim 0.025$\%
of the distance along a typical line of sight.  At present, only two
dozen debris disks have been spatially resolved.  Therefore it seems
unlikely that we have already imaged a debris disk residing in one of
these clouds.  Interaction with a cold, dense cloud is thus an
unsatisfactory explanation for the swept morphology observed in HD
61005.

\subsection{Interaction with a Warm, Low-Density Cloud}

Given that disk interaction with a cold, dense cloud appears unlikely,
we explore in this section whether interaction with a warm,
low-density cloud can potentially explain the observed morphology.

\subsubsection{Ram pressure stripping of disk gas and entrainment of disk grains}
\label{entrain}

Warm ($T \sim 7000 $ K) interstellar clouds dominate the mass of the
local interstellar medium (e.g., Frisch 2004) and occupy a local
volume filling factor of $\sim$5.5\%$-$19\% \citep{Redfield08}.  Our
optical spectrum does not constrain the presence of such a cloud
towards HD 61005, as typical columns towards these clouds are $N_{\rm
  H I} \sim 10^{17}$ cm$^{-2}$ \citep{Redfield06}.   

The low densities of warm local clouds are insufficient to supply a
ram pressure force comparable to the gravitational force and thereby
directly unbind grains from the system. For typical parameters of warm
local interstellar clouds ($n_{\rm H I}$ = 0.2 cm$^{-3}$, $v_{\rm
  rel}$ = 25 km s$^{-1}$; Redfield 2006) and 0.1 $\mu$m disk grains,
the ram pressure stripping radius is $\sim 10^{3}$ AU (Equation 1),
much larger than the radius of the observed bow structure ($\sim 2^{\prime\prime} =
69$ AU).  

However, in principle, ram pressure stripping could still play a role
if the HD 61005 disk contains gas that is undergoing ram pressure
stripping by the ISM. In this scenario---in contradistinction to the
direct ram pressure stripping scenario outlined in \S \ref{strip_bound}---disk grains
are swept away by the interstellar flow only because they are entrained in
outflowing disk gas. Gas-gas interactions correctly explain the HI
morphologies of galaxies travelling through an intracluster medium
(van Gorkom 2004); the truncated, swept-back HI disks of galaxies strongly
resemble HD 61005 (e.g., see Figures 1.7 and 1.8 of van Gorkom 2004).
We show below, however, that this interpretation for HD 61005 is
incorrect because the requirement on the density of disk gas is
incompatible with the requirement that grains be entrained.

To unbind disk gas, the ISM ram pressure must exceed the gravitational
force per unit disk area:
\begin{equation}
n \mu v^2 \gtrsim GM_{\rm star}\sigma / r^2
\end{equation}
where $\mu \approx 2 \times 10^{-24} \, {\rm g}$ is the mean molecular weight of
the ISM and $\sigma$ is the surface mass density of disk gas. For parameters
appropriate to a warm cloud \citep{Redfield06}, all disk gas having
a surface density
\begin{equation}
\begin{split}
\sigma \lesssim 4 \times 10^{-8} \left( \frac{r}{100 \, {\rm AU}} \right)^2 \left( \frac{0.95 \, M_{\odot}}{M_{\rm star}} \right) \left( \frac{n}{0.2 \, {\rm cm}^{-3}} \right) \\ \left( \frac{v}{25 \, {\rm km  \, s}^{-1}} \right)^2  \, {\rm g}\, {\rm cm}^{-2}
\end{split}
\end{equation}
is swept away by the ISM. The circumstellar gas content of HD 61005 is unknown.

For unbound disk gas to entrain disk grains, the momentum stopping
time of a grain in gas cannot be much longer than the outflow
timescale, $1/\Omega$, over which marginally unbound gas departs the
system, where $\Omega$ is the Kepler orbital frequency. From
Weidenschilling (1977), the momentum stopping time of a grain in
rarefied gas is given by the Epstein (free molecular drag) law as
\begin{equation}
\begin{split}
t_{\rm stop} \sim \frac{1}{\Omega} \frac{a_{\rm grain}\rho_{\rm grain}}{\sigma} \sim \frac{500}{\Omega} \left( \frac{a_{\rm grain}}{0.1 \, \mu{\rm m}} \right) \left( \frac{\rho_{\rm grain}}{2 \, {\rm g} \, {\rm cm}^{-3}} \right) \\ \left( \frac{4 \times 10^{-8} \, {\rm g} \, {\rm cm}^{-2}}{\sigma} \right)
\end{split}
\end{equation}
which is too long compared to the outflow time. Thus we discount the
possibility that the observed disk morphology arises from ram pressure
stripping of disk gas.

\subsubsection{Secular perturbations to grain orbits induced by ram pressure}
\label{secular}

The direct ram pressure stripping scenario described in
\S\ref{strip_bound} requires that the ISM ram pressure force on disk
grains be comparable to the stellar gravitational force.  As shown in
\S\ref{entrain}, this condition is not met in low-density warm clouds,
for which $F_{\rm ram} / F_{\rm grav} \sim 10^{-3}$ (at $\sim 70$ AU;
see Equation 1).  However, even in the case when the force exerted by
interstellar gas is much less than the gravitational force, neutral
gas can introduce secular perturbations to bound grain orbits that
could significantly change the morphology of the disk over timescales
of $\sim 10^3 - 10^4$ yr, assuming sub-micron sized grains dominate
the scattered-light distribution.  This perturbation timescale is less
than the crossing time of local warm clouds: $t_{\rm cross} \sim
(L_{\rm cloud} \,/ \, 5 {\rm \, pc}) / (v_{\rm rel}\, / \, 25 {\rm \,
  km\, s}^{-1}) \sim 10^5$ yr \citep{Redfield06}.  Thus this mechanism
can plausibly explain the disturbed HD 61005 morphology.
Interestingly, this scenario has been proposed as the primary removal
mechanism for dust in our own solar system at 20$-$100 AU
\citep{Scherer00}, though at present, little empirical evidence is
available to test this theory.

As described in \citet{Scherer00}, the underlying physical process
responsible for neutral gas drag is similar to that responsible for
solar wind drag.  In both cases, momentum transfer from incident
protons or gas particles to the grain surface results in secular
perturbations to the grain's initial orbit.  However, the
monodirectional character of the interstellar gas drag force leads to
changes in particle orbits that are very different from those induced
by the solar wind.  While the solar wind and Poynting-Robertson drag
both act to reduce grain eccentricities and semi-major axes,
interstellar gas drag tends to increase them, eventually unbinding the
grains from the system.  

\begin{figure}
\includegraphics[width=0.47\textwidth,angle=0]{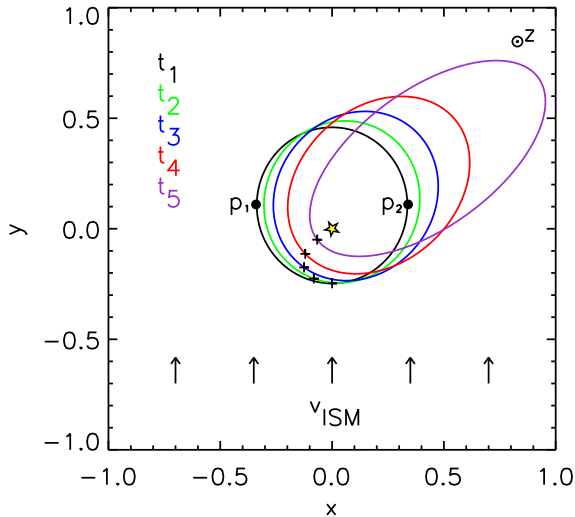} 
\caption{Orbital evolution of a test particle in the $x-y$ plane under
  the effect of neutral gas drag.  Five time periods separated by a
  fixed number of orbits are shown.  The grain begins with its
  star-pericenter (Runge-Lenz) vector anti-aligned with the incoming
  interstellar flow, directed along the positive $y-$axis, and its
  angular momentum vector aligned with the $z-$axis (out of the page).
  The ISM flow decelerates the grain on one leg (e.g., point $p_1$)
  and accelerates it on the other (e.g., point $p_2$), causing the
  Runge-Lenz vector to rotate.  The grain's periastron at each
  successive time is shown by a cross to illustrate this effect.}
\label{test_particles_peri}
\end{figure}

\begin{figure}
\includegraphics[width=0.57\textwidth,angle=0]{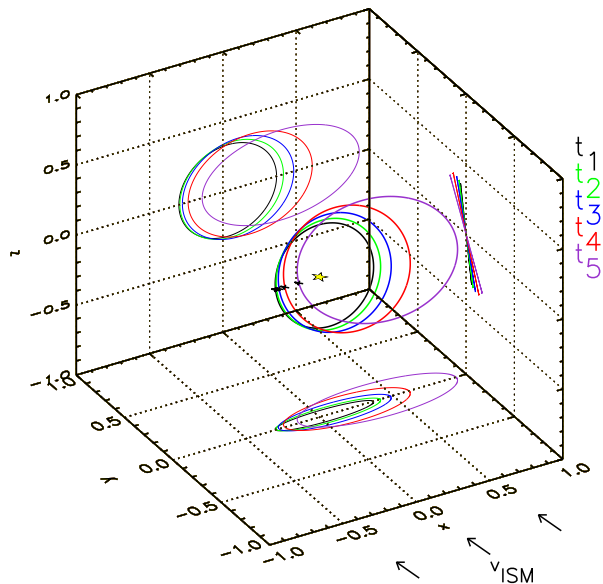} 
\caption{Orbital evolution due to neutral gas drag of a test particle
  initially inclined by $80^\circ$ with respect to the incoming flow.
  Five time periods separated by a fixed number of orbits are shown.
  The grain's initial Runge-Lenz vector is chosen to be at its
  equilibrium orientation. The ISM flow exerts a net torque on the
  orbiting grain, causing the angular momentum vector to rotate into a
  direction perpendicular to the ISM flow.}
\label{test_particles_orbplane}
\end{figure}

In the absence of other perturbing forces, the analytic work of
\citet{Scherer00} shows that the gas drag force acts to rotate a given
particle's orbital plane into a plane coplanar to the flow vector, and
its star-pericenter (Runge-Lenz) vector into a direction perpendicular
to the flow vector.  These effects are illustrated in Figures
\ref{test_particles_peri} and \ref{test_particles_orbplane}, which
show the orbital evolution of two dust grains with different initial
orbital elements.  The grain in Figure \ref{test_particles_peri}
starts on a low-eccentricity ($e=0.3$) orbit in the $x-y$ plane with
its Runge-Lenz vector anti-aligned with the incoming interstellar flow
and its angular momentum vector aligned with the $z-$axis (out of the
page). This orientation causes the grain to be decelerated on one leg
(e.g., point $p_1$) and accelerated on the other (e.g., point $p_2$),
causing the Runge-Lenz vector to rotate.  The grain's periastron at
each successive time is denoted by a cross to highlight this effect.
The rotation continues until the Runge-Lenz vector becomes
perpendicular to the incoming flow.  Thus, counterintuitively, neutral
gas drag leads to a build up of grains perpendicular to the relative
flow direction.

The tendency of neutral gas drag to rotate a given grain's orbital
plane into a direction coplanar with the incoming flow is illustrated
in Figure \ref{test_particles_orbplane}, which shows the evolution of
a low-eccentricity ($e=0.1$) grain initially inclined $80^\circ$ with
respect to the ISM flow.  The grain's initial Runge-Lenz vector is
chosen to be at its equilibrium orientation (perpendicular to the flow)
to isolate the effect of the orbital plane rotation.  The ISM flow
exerts a net torque on the orbiting grain, causing the angular
momentum vector to rotate into a direction perpendicular to the
incoming flow.  This effect is easiest to discern from the
two-dimensional projection of the $y-z$ plane in Figure
\ref{test_particles_orbplane}.

The above discussion suggests that rigorously modeling the neutral
gas drag effect requires knowledge of the initial grain orbital
elements and the interstellar gas flow.  However, for the case of HD
61005, neither of these prerequisites is known. HD 61005 has a
well-determined space motion; the Hipparcos-measured proper motion
corresponds to a plane-of-sky velocity of $v_\alpha = 9.2 \pm 0.3$ km
s$^{-1}$, $v_\delta = 12.2 \pm 0.4$ km s$^{-1}$ at the distance of HD
61005.  The radial velocity is $v_r = 22.3 \pm 0.2$ km s$^{-1}$
\citep{Nordstrom04}.  However, the velocity of the putative cloud
responsible for the swept morphology is unknown.  Velocities of local
warm clouds can be comparable to the observed space motion of HD 61005
\citep{Frisch02}.  Thus the red arrow in Figure 3 of H07 denoting the
direction of the star's tangential motion is not a reliable indicator
of the cloud-star relative velocity.

Unfortunately, HD 61005 is difficult to assign to any known
interstellar clouds, owing to its heliocentric distance (34.5 pc) and
galactic coordinates ($l = 246.4 ^\circ, b = -5.6 ^\circ$).
\citet{Redfield08} recently used radial velocity measurements of 157
lines of sight to identify 15 warm clouds in the local interstellar
medium.  However, the identified clouds are thought to reside largely
within 15 pc of the Sun, whereas the distance to HD 61005 is 34.5
pc. Furthermore, the line of sight to HD 61005 is not assigned to any
of these clouds. The star's galactic coordinates could plausibly
associate it with either the G cloud or the Blue cloud (e.g., see
Figure 19 in Redfield \& Linsky 2008), though HD 61005 is more likely
associated with a more distant, currently unidentified cloud,
rendering the cloud-star relative velocity for this system highly
uncertain.

The initial orbital elements prior to the presumed cloud interaction
are similarly unknown.  The radial architectures of debris disks in
scattered light show significant diversity and currently appear
largely independent of other observables, such as age or spectral type
\citep{Kalas06}.  Furthermore, as discussed in \S \ref{sbp_section},
the one-dimensional surface brightness profiles of HD 61005 do not
allow us to place strong constraints on the location of the parent body
population, and thus the grain eccentricities and inclinations.  We do
note, however, that the sub-micron particle sizes implied by the
polarization and color measurements do not {\it a priori} contradict
the bound orbits requirement of the gas drag perturbation theory, as
radiation pressure from the star is likely insufficient to unbind
grains of any size (Appendix A).

Given the large uncertainties in the disk and ambient ISM properties
of the HD 61005 system, we are unable to empirically test whether gas
drag is responsible for the asymmetric, swept morphology.  Therefore
to explore whether this mechanism can plausibly explain the observed
structure, we adopt the numerical techniques described in
\citet{Scherer00} to construct a modest grid of models with reasonable
assumptions for the disk grains and ISM.  Details of the model
construction and restrictions are described in Appendix B.  In
summary, we subject a ring of 0.1 $\mu$m, low eccentricity ($e=0.2$)
grains with semimajor axes of 60 AU and random inclinations to a
uniform density cloud typical of nearby warm interstellar clouds
($n_{\rm HI} = 0.2$ cm$^{-3}$) traveling at a typical cloud-star
relative velocity of $v_{\rm rel} = 25$ km s$^{-1}$
\citep{Redfield06}.  We test a range of relative flow vectors and
produce scattered light images from the resultant grain distributions
after the system has achieved a steady state.

\begin{figure*}
\begin{center}
\includegraphics[width=0.75\textwidth,angle=0]{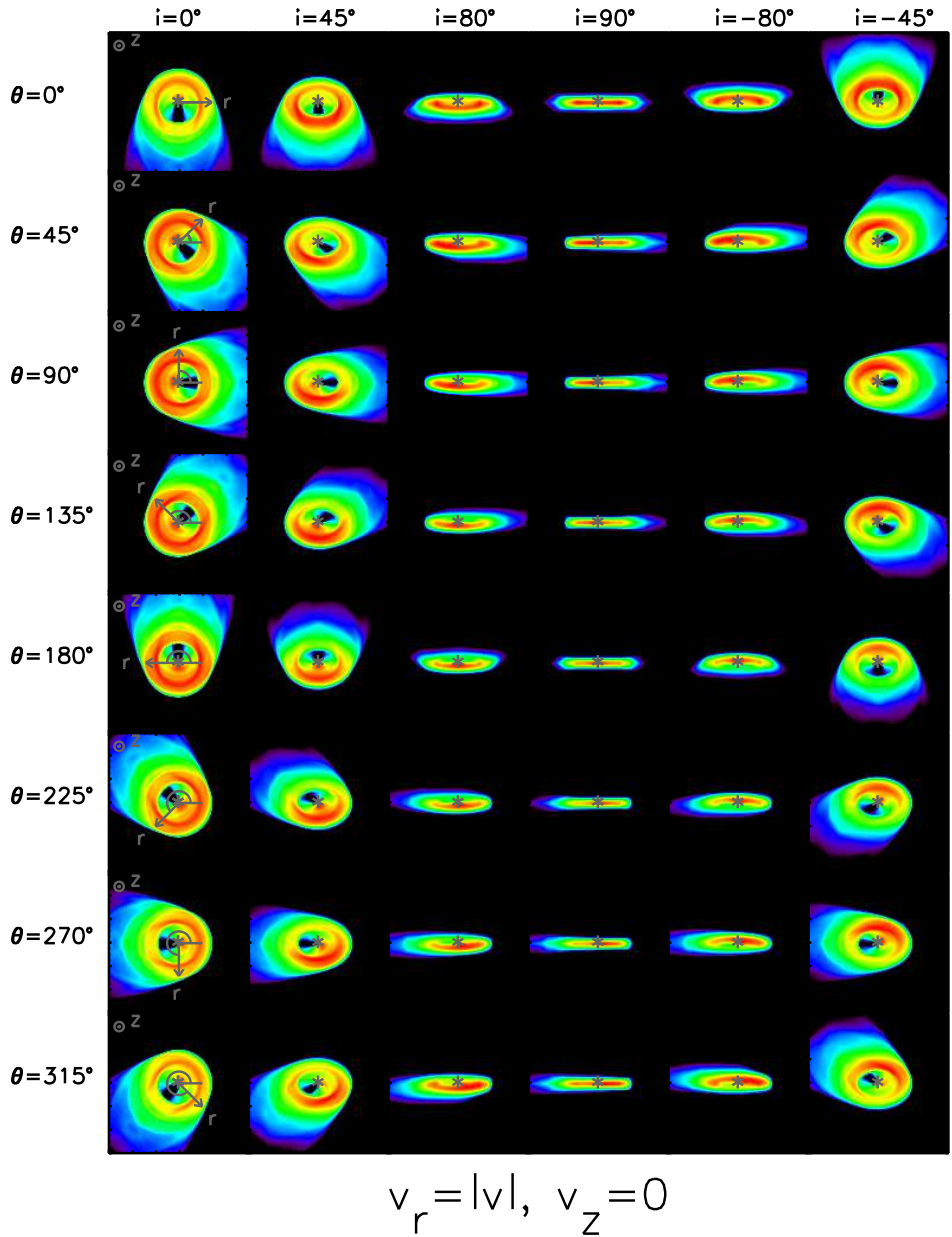}
\end{center}
\caption{Steady state model images for hypothetical debris disk
  systems undergoing neutral gas perturbations.  In all models, the
  relative flow is coplanar with the disk midplane.  The adopted
  cylindrical coordinate system is shown with respect to the face-on
  disks in the left column.  The vector $\boldsymbol{r}$ points in the
  direction of the relative ISM flow; the azimuthal orientation of
  $\boldsymbol{r}$ is defined by $\theta$.  The disk inclinations are
  indicated at top.  Each box is $9^{\prime\prime} \times
  9^{\prime\prime}$ (assuming a distance to the system of 34.5 pc);
  the color scale is logarithmic. The models show that brightness
  asymmetries, bow structures, and swept morphologies can all be
  produced by disk encounters with warm interstellar clouds, which
  occupy a sizable fraction of the local ISM. \vspace*{3mm}}
\label{vr}
\end{figure*}

\begin{figure*}
\begin{center}
\includegraphics[width=0.75\textwidth,angle=0]{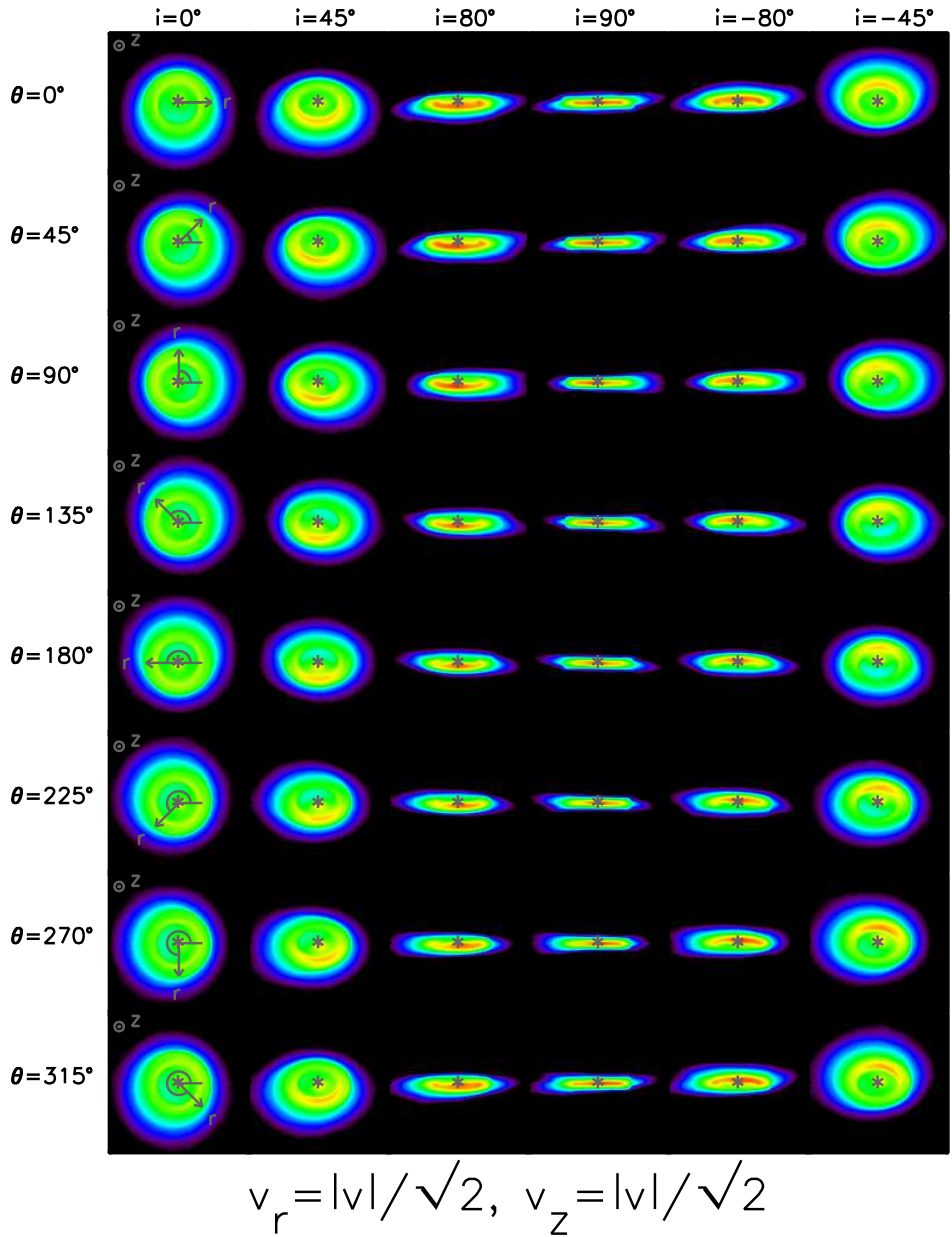}
\end{center}
\caption{Same as Figure \ref{vr}, only with components of the relative
  flow both coplanar with the disk midplane and perpendicular to it.  The
  radial and perpendicular flow components are equal in magnitude.}
\label{vrvz}
\end{figure*}

\begin{figure*}
\begin{center}
\includegraphics[width=0.75\textwidth,angle=0]{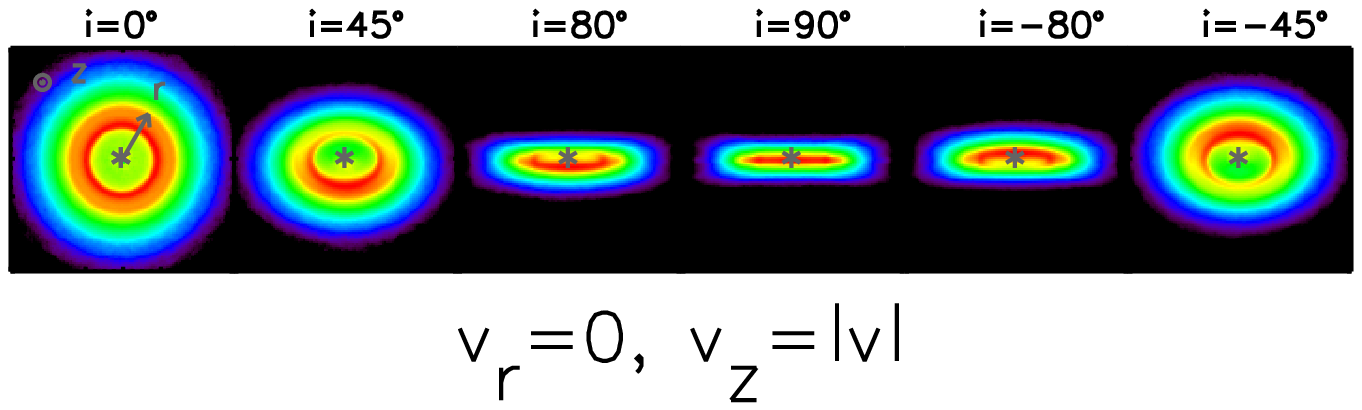}
\end{center}
\caption{Same as Figures \ref{vr} and \ref{vrvz}, only with the
  relative flow vector oriented purely perpendicular to the disk
  midplane.  The dust distribution in these models is largely
  axisymmetric, as described in the text.  The brightness asymmetries
  evident at intermediate inclinations
  ($i=45^\circ,80^\circ,-80^\circ,-45^\circ$) are the result of
  scattering asymmetry; for positive inclinations, the lower half of
  the disk in the image is closer to the observer than the upper
  half. \vspace*{3mm}}
\label{vz}
\end{figure*}

The resulting grid of models produced for a range of relative flow
directions and disk inclinations is presented in Figures \ref{vr} $-$
\ref {vz}.  Each frame is $9^{\prime\prime}$ across, and the color
scale is logarithmic.  The adopted coordinate system is described in
Appendix B.  A comparison of Figure \ref{acs_image_log} to Figures
\ref{vr} $-$ \ref {vz} suggests that none of the IS gas drag models
are a striking match to the data.  On the other hand, a
gross swept morphology, somewhat similar to HD 61005, is present in a
few of the model panels.  These best approximations correspond to disk
inclinations of $i = \pm80^\circ$, consistent with the data (\S
3.1.2), and relative flow directions largely coplanar with the disk
midplane, approximately in the plane of the sky (Figure \ref{vr},
panels $\theta \sim 0^\circ, 180^\circ$).  This relative flow
direction is counterintuitive, going against the flow direction
suggested by preliminary inspection of the scattered-light images,
perhaps providing a further strike the against IS gas drag
interpretation.  On the other hand, if these preliminary models are a
first approximation of the data, then this distinctive relative
velocity may be verified by future observations (\S 5.1).  

\begin{figure}
\includegraphics[width=0.47\textwidth,angle=0]{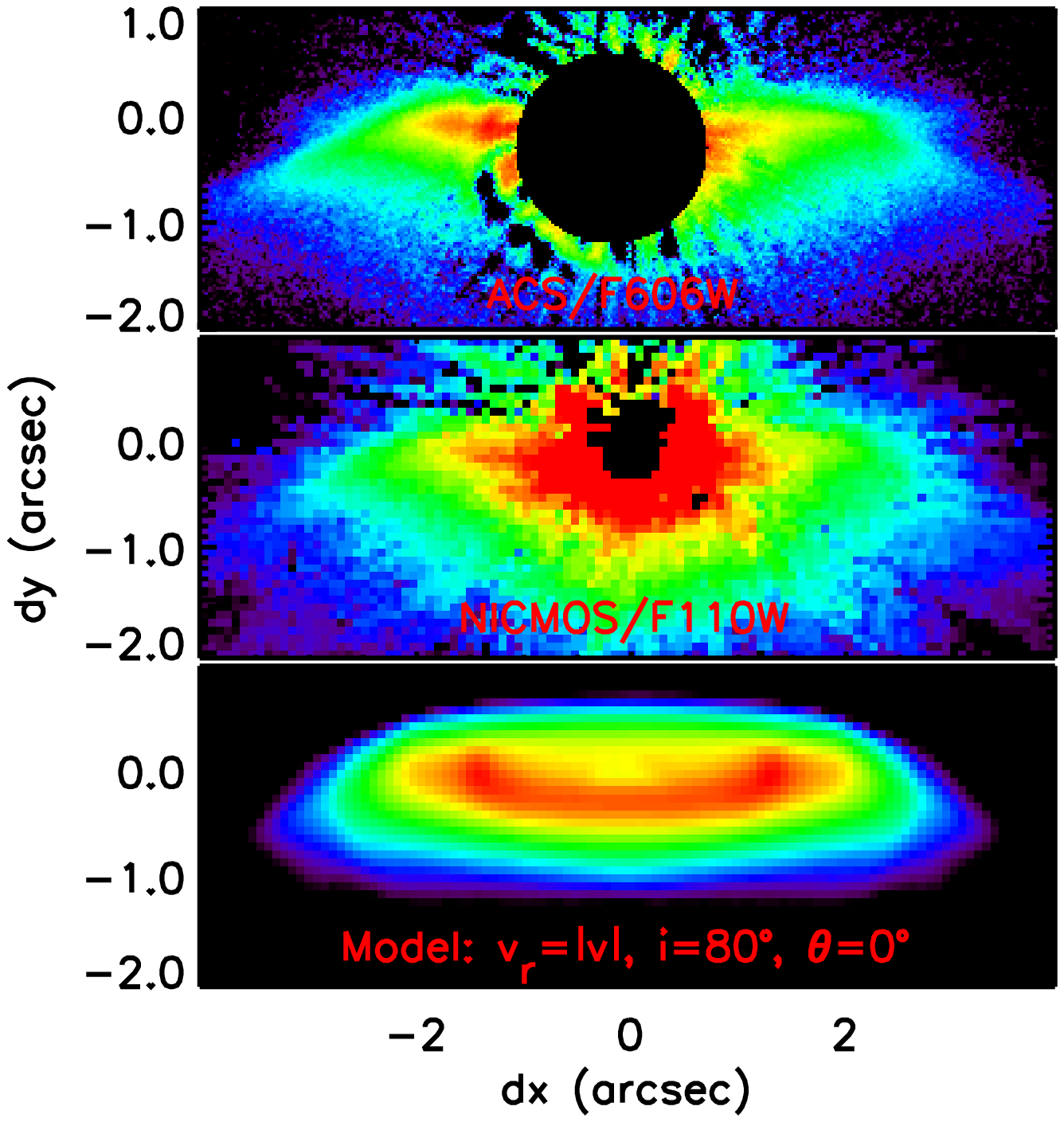} 
\caption{Comparison between the ACS Stokes $I$ image (top), the NICMOS
  1.1 $\mu$m image (middle), and a promising neutral gas model
  (bottom), taken from Figure \ref{vr} (top row, third column).  This
  comparison shows that the simple model presented here is only a very
  rough representation of the data.  Still, the gross swept,
  asymmetric morphology is clearly present in the model image.\vspace*{3mm}}
\label{acs_nicmos_model}
\end{figure}

For ease of viewing, we show in Figure \ref{acs_nicmos_model} the ACS
and NICMOS data together with a neutral gas drag model that roughly
approximates the observed morphology, corresponding to the top row,
third panel of Figure \ref{vr}.  A summary of the morphological
shortcomings of this strongest IS gas drag model can be summarized as
follows:
\begin{enumerate}
\item{The swept back components NE2 and SW2 are much more extended
    relative to the main disk in the data than in the model.}
\item{The edges of the swept back structure are more pronounced 
    in the data than in the model.}
\item{The sharp radial spurs observed beyond the inflection point
    along components NE1 and SW1 are not present in the model.}
\item{The model figure does not show the significant brightness
    asymmetry between the northeast and southwest disk lobes that is
    seen in the data.}
\end{enumerate}
These differences could indicate that the physics incorporated
into the current IS gas drag models is overly simplistic (\S 5.1), or
that interstellar gas drag is not responsible for the observed
morphology (e.g., \S 5.3). 

As discussed in \S 3.1, the ACS polarization results can be understood
qualitatively through geometric considerations alone.  Thus the
most promising IS gas drag models naturally reproduce the observed
polarization structure.

\section{Discussion}

\subsection{Interstellar Gas and the HD 61005 Morphology}

The previous section explored whether disk/gas interaction can
plausibly explain the unusual HD 61005 morphology.  Of the four
scenarios considered, three are implausible, given the limits on the
ambient interstellar gas density imposed by our optical spectrum.  The
fourth scenario, secular perturbations from low density gas, is an
attractive alternative, as this mechanism can significantly distort
grain orbits well within a cloud crossing time.  Furthermore, the
densities required by this scenario are typical of local interstellar
clouds, which occupy up to $\sim20$\% of the local ISM.

Nevertheless, our preliminary modeling of this effect (Figures
\ref{vr} $-$ \ref {vz}) can only produce disk morphologies in very
rough agreement with the observations, suggesting that either
additional physics needs to be incorporated into the current models,
or that an altogether distinct physical mechanism is at work.  Indeed,
the current models are simplistic, and their applicability is limited
by several key assumptions:

\begin{enumerate}
\item{{\it Astrosphere Sizes:} As discussed by \citet{Scherer00}, the
    toy models presented in Figure \ref{vr} $-$ \ref {vz} require that
    the disturbed grains be inside the astrospheric termination shock,
    such that the interstellar gas density and velocity can be
    approximated as constant. For the case of HD 61005, the
    termination shock distance is unknown. Furthermore, as HD 61005 is
    farther away than any star for which a direct astrospheric
    detection has been made \citep{Wood04}, the astrosphere size may
    be difficult to constrain observationally.  In general,
    termination shock distances vary greatly, depending on the ambient
    ISM and stellar wind conditions (e.g., densities, temperatures,
    velocities, stellar activity).  For the case of the Sun only, the
    hydrodynamic models of \citet{Muller06} show that the termination
    shock distance could easily vary between $\sim 10$ AU and $500$
    AU.  Observational astrosphere measurements of solar-type stars
    are consistent with these predictions\footnote{ On a broader
      scale, no astrosphere detections have been made for stars
      earlier than G-type.  As a result, the typical effect of
      astrospheres on IS gas drag in debris disks surrounding A-type
      stars is difficult to reliably assess.} (Mann et al. 2006, and
    references therein). }
\item{{\it Initial Conditions:} As noted in \S 4.2.2 and Appendix B,
    our adopted initial orbital elements for the HD 61005 disk grains
    prior to the interstellar cloud interaction are highly uncertain,
    given a lack of information for the grain properties and
    underlying planetesimal population that collisionally replenishes
    the observed dust disk. Future long-wavelength observations
    sensitive to larger grains may be able to place tighter
    constraints on the distribution of sub-micron grains prior to the
    interstellar cloud interaction, as the distribution of large
    grains would likely reflect that of the parent bodies for the
    sub-micron population.  Furthermore, larger grains would not be
    significantly affected by interstellar gas drag on the same
    timescale as the sub-micron size grains traced in these
    observations.  The numerical techniques employed here and in
    \citet{Scherer00} can be easily revised to accommodate an
    arbitrary initial disk architecture, provided the orbits are not
    highly eccentric, such that averaging over one orbit and applying
    Gauss' method is invalid.}
\item{{\it Internal Disk Collisions:} In Appendix B, we estimate to
    order-of-magnitude that the collision time for submicron grains at
    $\sim$70 AU is $\sim$5000 yr. This collision time is somewhat
    longer than the timescale over which our model relaxes to a steady
    state --- essentially the time for gas drag to unbind a grain ---
    given in Appendix B as $\sim$3000 yr. That the times are
    comparable supports the assumption of our models that each grain
    removed by gas drag is collisionally replenished. At the same
    time, the comparison of timescales underscores a shortcoming of
    our model --- that removal of grains by collisions is ignored. In
    reality, submicron grains should be removed from the system not
    only by gas drag, but also by collisions, in roughly equal
    proportions.  We defer to future work a comprehensive study that
    includes removal by collisions via a collisional cascade.}
\item{{\it Planetary Configurations:} Finally, as illustrated by
    \citet{Scherer00} (e.g., his Figure 2), the incorporation of
    planetary orbits can appreciably change the perturbed orbital
    elements from the case in which only IS gas drag is considered.
    This caveat is especially important for massive grains, or grains
    in close proximity to planetary orbits.  As a result, the models
    presented here should be treated with some caution if applied to
    typical planetary system scales ($\lesssim 50$ AU; e.g., Kenyon \&
    Bromley 2004).  This cautionary point may be particularly
    important for the case of HD 61005, as the origin of the
    brightness asymmetry between the northeast and southwest disk
    lobes (\S 3.1.1) is unknown.  The agreement between the northeast
    and southwest deflected component position angles (\S 3.1.3 and
    Figures \ref{acs_image_log} $-$ \ref{acs_image_lin}) suggests this
    asymmetry may originate from a physical mechanism entirely
    distinct from ISM interaction.  If the asymmetry is due to a
    massive perturber, the disk morphologies produced in Figures
    \ref{vr} $-$ \ref{vz} are likely to be inapplicable.  Resolved
    long wavelength observations sensitive to massive grains are
    needed to further explore this possibility.}
\end{enumerate}

In addition to the above uncertainties, a remaining ambiguity
important for future IS gas drag modeling is the velocity of the
putative cloud responsible for the HD 61005 morphology.  H07 noted
that the star's tangential space motion is perpendicular to the disk
midplane, in agreement with the relative flow vector suggested by
initial inspection of the observed images.  However, while assigning
the relative flow direction to the star's tangential velocity is
appealing, the three scenarios explored in \S 4 which assume this flow
direction were found to be untenable.  Furthermore, velocities of
local warm clouds can be comparable to the observed space motion of HD
61005 (\S 4.2.2).  As a result, the star's tangential motion is not a
reliable indicator of the cloud-star relative velocity.

Future spectroscopic observations may be able to detect the cloud
directly (e.g., HST/GO Program 11674; H. Maness, PI), thereby providing
key constraints on the ambient ISM density and velocity.  Such
observations will greatly inform future modeling, as the preliminary
interstellar gas drag models presented here suggest a counterintuitive
relative motion parallel to the disk midplane, rather than
perpendicular to it.  Though HD 61005 is difficult to assign to any
known interstellar clouds (\S 4.2.2), its galactic coordinates could
plausibly associate it with either the G cloud or the Blue cloud.  The
space velocities of both of these clouds suggest a relative motion
dominated by the radial velocity component.  Therefore if HD 61005 is
associated with either of these clouds, the relative motion is
inconsistent with all models posited in \S 4.

\subsection{General Applicability of Interstellar Gas Drag}

The normalcy of the interstellar densities, velocities, and cloud
sizes required by the secular perturbation model in \S 4.2.2 suggests
that IS gas drag can be important beyond HD 61005 in shaping debris
disk morphologies.  Taking the simple models of \S 4.2.2 and Appendix
B at face value, several of the general morphological features
produced in Figures \ref{vr} $-$ \ref{vz} are consistent with observed
disk structures.  For example, the extreme brightness asymmetry in HD
15115 \citep{Kalas07} may potentially result from interstellar gas
drag, though a range of alternative explanations could explain this
system as well (e.g., see list in \S 1).  The bow structures seen in
some of the face-on models in Figure \ref{vr} are also reminiscent of the
mid-infrared morphology observed around the A star, $\delta$ Velorum,
which was recently modeled as a purely interstellar dust phenomenon
\citep{Gaspar08}.  Finally, the middle panels in Figure \ref{vrvz}
show that warps, similar to that seen in $\beta$ Pic (e.g., Mouillet
et al. 1997), can in principle be produced for a relatively wide range
of flow directions.

However, while IS gas drag can in principle produce commonly observed
disk features, the rate at which gas drag is expected to affect the
observations remains unclear. Beyond the uncertainties in the model
physics described in \S 5.1, the characteristics of warm, low density
clouds are currently uncertain, as detailed knowledge of them is
limited to clouds residing predominantly within 15 pc of the Sun
\citep{Redfield08}.  As a result, our understanding of typical cloud
sizes, shapes, and total volumetric filling factor remains
rudimentary.  A key finding in this area, however, is that a
significant fraction of nearby warm clouds appear to exhibit
filamentary morphologies, which would limit the average interaction
time between a given disk and cloud, likely reducing the rate at which
IS gas perturbations produce an observable effect.  This concern is
particularly important for the case of disks surrounding early-type
stars, as grains traced in scattered light tend to be larger in disks
surrounding A-type stars than in their later type counterparts, owing
to the larger radiation pressure blowout size.  As such, the
scattered-light morphologies for A-star disks require correspondingly
longer cloud-disk interaction times to be noticeably affected.  The
timescale for a given grain to become unbound under IS gas drag
increases approximately as the square root of the grain size
\citep{Scherer00}.

\subsection{Interstellar Grains and the HD 61005 Morphology}

Finally, we note that all models posited in \S 4 consider only the
role of interstellar gas, ignoring the potential effects of
interstellar grains. \citet{Artymowicz97} investigated IS
sandblasting of debris disks surrounding A stars and found that
sandblasting has a negligible effect on the observed structure, as
radiation pressure blows out most incoming interstellar grains before
they are allowed to intersect the disk.  However, under this
framework, only grains with $\beta \geq 1$ are ejected.  Thus Figure
\ref{radpr} shows that radiation pressure does not protect the HD
61005 disk, as it does in A-stars.

Nevertheless, even if radiation pressure does not protect the disk
against sandblasting, the stellar wind might, as only large
interstellar grains with sizes greater than a few $\times 0.1 \mu$m
are allowed to enter astrospheres freely without deflection
\citep{Linde00}.  Thus it is likely that interstellar sandblasting can
only plausibly compete with interstellar gas drag if the astrosphere
is smaller than or comparable to the observed debris disk size
\citep{Mann06}.  The size of the HD 61005 astrosphere is unconstrained
by present observations.  In general, observations and models of
astrospheres surrounding solar-type stars show sizes in the range
$\sim 10-10^3$ AU, depending on the ambient ISM and stellar wind
conditions (see discussion in \S 5.1).  Thus with a characteristic
disk size of $\lesssim 70$ AU, it is not clear whether typical
interstellar grains can intersect the HD 61005 disk.

Detailed modeling of sandblasting is outside the scope of this paper.
However, future theoretical work should investigate the effects of
sandblasting on debris disks surrounding solar-type stars.
Calculations of the ISM density required for sandblasting to
eject an observable flux of grains, the disk morphologies produced in
this case, and the timescale for which sandblasting can be sustained
would significantly aid in differentiating between this explanation
and the gas drag models presented here.

\section{Summary}

The morphology and polarization structure of HD 61005 in the HST/ACS
data (Figures \ref{acs_image_log} $-$ \ref{moth_ip}) strongly suggest
that HD 61005 is a debris disk undergoing significant erosion by the
ambient interstellar medium.  The physical mechanism responsible for
this erosion remains uncertain.  Previous work has suggested that HD
61005 may be interacting with an unusually dense cloud.  However, our
high-resolution optical spectrum argues against this idea, instead
suggesting an ambient ISM density typical of local interstellar
clouds.  Thus the evolutionary state of HD 61005 may represent a
commonplace, intermittent stage of debris disk evolution driven by
interaction with typical, low-density gas.

With this motivation, we considered the effects of secular
perturbations to grain orbits induced by ram pressure in warm, tenuous
clouds.  This mechanism can significantly distort grain orbits within
a typical cloud crossing time and generate structures that very
roughly reproduce the HD 61005 images.  Future work that incorporates
additional, more detailed physics may improve the agreement between
the observations and interstellar gas drag models. The theoretical
effects of interstellar sandblasting for solar-type stars should also
be investigated in greater detail.

Regardless of the interpretation for HD 61005, we expect interstellar
gas drag is important at some level in shaping the structure and
evolution of planetary debris disks.  The frequency with which this
effect is important strongly depends on the typical sizes, shapes,
velocities, and filling factors of warm interstellar clouds, which
have poorly constrained global properties at present.  Nevertheless,
some morphological features common to nearby resolved debris disks
(e.g., brightness asymmetries, warps, and bow structures) can in
principle be produced in this way.  A larger sample of spatially
resolved debris disks at a wide range of wavelengths and more detailed
theoretical work will help eliminate some of these remaining
ambiguities.

\acknowledgments We wish to thank Jay Anderson and Vera Platais for
providing the astrometric software used to test for companionship.  We
also thank Carl Heiles, Gaspard Duchene, Seth Redfield, and Marshall
Perrin for useful conversations that helped shape the ideas discussed
in this paper.  H.M. is funded by the GRFP at NSF and the GOPF at UC
Berkeley. Support for this work was provided by NASA through grant
number GO-10847 from the Space Telescope Science Institute, which is
operated by Association of Universities for Research in Astronomy
Incorporated, under NASA contract NAS5-26555.  This work was also
supported in part by the University of California Lab Research Program
09-LR-01-118057-GRAJ and NSF AST-0909188.

\appendix

\section{Radiation Pressure and Blow-out}

To provide a preliminary assessment of the unbound grain contribution,
we compute the radiation pressure force according to
\citet{Kruegel03}\footnote{This expression is equivalent to that given by \citet{Kohler04}: \\ $F_{\rm RP} = \int \frac{\pi
      a^2}{c} [Q^{\rm{abs}}_\nu + (1-{\rm g}_\nu) Q^{\rm{sca}}_\nu] F_\nu \;d\nu.$}:
\begin{equation}
F_{\rm RP} =   \int \frac{\pi a^2}{c} (1-{\rm g}_\nu \omega_\nu ) Q^{\rm{ext}}_\nu F_\nu \;d\nu.
\label{eq-rad-press}
\end{equation} 
Here, g$_\nu$ is the grain scattering asymmetry, $\omega_\nu$ is the
albedo, and $Q^{ext}_\nu$ is extinction cross section in units of the
geometric cross section. H07 and \citet{Carpenter08} showed that the
stellar spectral energy distribution of HD 61005 is well-matched by a
main-sequence $T_{\rm eff} = 5456$ K Kurucz model atmosphere.  We
therefore use their best-fit spectrum in evaluating
Eq. (\ref{eq-rad-press}) to interpolate between the data and to
extrapolate for the small fraction of missing flux longward of
24 $\mu$m and shortward of 0.3 $\mu$m.

\begin{figure}
\begin{center}
\includegraphics[width=0.7\textwidth,angle=0]{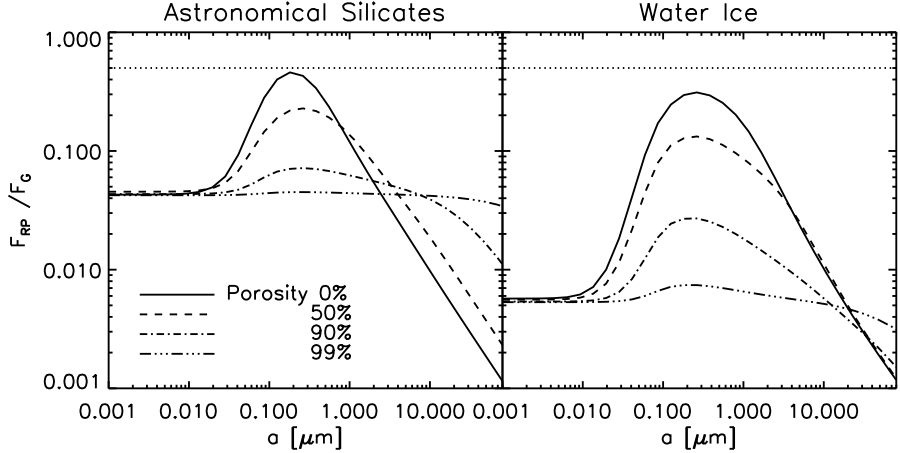} 
\end{center}
\caption{Ratio of the radiation pressure force to gravity ($\beta$)
  for astrophysical silicate grains (left) and water ice (right) for
  the HD 61005 system.  The horizontal dotted line indicates the ratio
  above which grains launched by parent bodies on circular orbits
  become unbound. Thus silicate and ice grains in a conventional
  debris disk surrounding a solar type star are likely to remain bound
  to the star.  On the other hand, the weak radiation field ($\beta <
  1$) implies that radiation pressure alone does not impede
  interstellar grains from entering the system.  Thus if the
  astrosphere surrounding HD 61005 is smaller than the disk,
  interstellar sandblasting could potentially erode the HD 61005 disk
  grains.}
\label{radpr}
\end{figure}

Figure \ref{radpr} shows the ratio of the radiation pressure to
gravitational force, $\beta = F_{\rm RP}/F_{\rm G} $, for spherical
particles with scattering properties computed using Mie
theory. Results are shown for water ice ($\rho = 1$ g cm$^{-3}$;
Warren 1984) and Draine's astrophysical silicate ($\rho = 3.5$ g
cm$^{-3}$; Draine \& Lee 1984).  Mie theory radiation pressure
calculations for silicate grains have been verified to within factors
of a few using results from microwave analog laboratory data,
computational Discrete Dipole Approximation (DDA) and T-matrix
calculations, and solar system collection experiments (Wehry et
al. 2004, Landgraf et al. 1999, and references therein).  Solar system
calculations performed on non-spherical, ballistic particle-cluster
aggregates and ballistic cluster-cluster aggregates also yield similar
results to spherical-grain Mie theory calculations (e.g., Fig. 7 Mann et
al. 2006, and references therein).  For the various porosities shown
in Figure \ref{radpr}, we used the Maxwell-Garnett rule to compute the
approximate dielectric constant for a dilute medium.  As applied to
debris disk systems, this method has been found to agree well with DDA
calculations for aggregate porosities of $P \lesssim 90$\%
\citep{Kohler04}.

The calculation in Figure \ref{radpr} does not include the effect of
stellar wind pressure, as the stellar wind parameters for HD 61005 are
unknown, and the X-ray flux of HD 61005 ($F_X = 4 \times 10^6$;
Wichmann et al. 2003) exceeds the maximum value for which the
$F_X-\dot{M}_*$ relation of \citet{Wood05} is reliable.  Nevertheless,
evaluating Equation 7 from \citet{Strubbe06} in the geometric optics
limit, stellar wind pressure is predicted to be over an order of
magnitude less significant than radiation pressure, even for mass loss
rates approaching $\dot{M}_* \sim 100\, \dot{M}_\odot $, approximately
the highest mass loss rate yet observed.  Thus the radiation pressure
force of Figure \ref{radpr} is expected to be representative of the
total outward pressure force, independent of the stellar mass loss
rate.

If a grain is liberated from a parent body that is on a circular
orbit, radiation pressure increases its eccentricity such that $e =
\beta/(1-\beta)$; for $\beta \geq 1/2$, the grain is launched onto an
unbound orbit.  Debris disk grains tend to be highly porous with
vacuum volume filling fractions of $\sim 90\%$ (e.g., Li \& Greenberg
1998, Graham et al. 2007).  Thus Figure \ref{radpr} shows that
silicate and ice grains in a conventional debris disk surrounding a
late-G star like HD 61005 are likely to remain bound to the star,
unless the parent body population is highly eccentric.

\section{Neutral Gas Drag Toy Models}

\subsection{Model Construction and Results}

The time for a 0.1 $\mu$m grain to become unbound by interstellar gas
drag is $\sim 10^3 - 10^4$ years, depending on the ambient ISM
density, velocity, and relative flow direction, and the initial
orbital elements of the grain (more precise estimates are given later
in this section).  This timescale may be compared to the collision
time for submicron grains.  Low eccentricity ($e =
0.1-0.2$) grains are pumped to moderate eccentricities ($e=0.3-0.4$)
by interstellar gas drag within a typical orbital period of $500$ yr.
Thus, combining equations 7 and 9 from \citet{Chiang09} for moderately
eccentric grains, the approximate collisional lifetime of submicron
sized grains is

\begin{equation}
t_{\rm col} \sim 5000\, {\rm yr}\, 
\frac{2 \times 10^{-3}} {L_{\rm IR}/L_{*}}
\frac{H/R}{0.1} 
\sqrt{\frac{\Delta R/R}{0.2}} 
\left( \frac{0.95\,M_\sun}{M_*} \right) ^{1/2}
\left(  
\frac{R}{70\, \rm{AU}} \right) ^{3/2}
\left( \frac{1-0.4}{1-e} \right)^{3/2} .
\label{tcol}
\end{equation}

Here, the fractional luminosity, ${L_{\rm IR}/L_{*}} = 2 \times
10^{-3}$, is taken from H07 and is identical to the result obtained
from integrating the best fit SED in \citet{Roccatagliata09}.  Their
 fit implies that grains smaller than $\sim 1 \,\mu$m are
responsible for the bulk of the infrared excess and that these grains
reside at a minimum distance of $96 \pm 23$ AU from the central star.
This minimum disk radius is consistent with the characteristic radius
of 70 AU adopted in Equation \ref{tcol}.


Since the collision time appears comparable to the time for the ISM to
remove grains, we assume for our numerical models that a steady state
is established in which the collisional generation of submicron grains
within the posited birth ring is balanced by their removal by
interstellar gas drag. We neglect the depletion of grains by
collisions---this amounts to an order-unity error (see also Section
5.1, item 3). Within this framework, we follow the evolution of $10^5$
0.1 $\mu$m particles.  This single grain size is chosen for simplicity
and is meant to be representative, given the polarization and color
results described in \S 3.  Based on the observed morphology, we
choose an initial semimajor axis for all grains of 60 AU.  Assuming
the observed grains are collisionally created by grains on circular
orbits, Appendix A suggests the initial grain eccentricities can range
from $e \sim 0$ to $e \sim 0.4$ for 0.1 $\mu$m grains, depending on
the grain material and porosity.  We adopt a representative value of
$e=0.2$ in our models, appropriate for moderately porous silicate
grains.  This eccentricity is consistent with the fractional ring
width adopted in Equation \ref{tcol}.  We further choose random
inclinations drawn from a uniform distribution extending from
0$^\circ$ to 10$^\circ$, based on an assumed disk aspect ratio of $H/R
\sim 0.1$, consistent with the system morphology in Figures
\ref{acs_image_log} and \ref{acs_image_lin} and with measurements of
other highly inclined debris disks \citep{Golimowski06, Kalas05,
  Krist05}.  Finally, we assume that the disk is initially circularly
symmetric, such that the longitudes of ascending nodes, mean
anomalies, and arguments of periastra are uniformly distributed
between $0^\circ$ and $360^\circ$.

Having chosen initial orbital elements, we next calculate the secular
perturbations to the angular momentum and Runge-Lenz vectors from
interstellar gas drag by averaging the Gaussian perturbation equations
numerically over one orbit \citep{Brouwer61}.  We adopt an ISM density
and encounter speed typical of warm interstellar clouds ($n_{\rm H I}$
= 0.2 cm$^{-3}$, $v_{\rm rel}$ = 25 km s$^{-1}$; Redfield 2006),
though we note that the gas density is likely reduced by a factor a
few inside the astrosphere relative to the nominal value outside
\citep{Bzowski09}. The direction of the flow vector relative to the
disk has a significant effect on the resulting disk morphology, and we
therefore test a range of flow directions.

\begin{figure}
\begin{center}
\includegraphics[width=0.6\textwidth,angle=0]{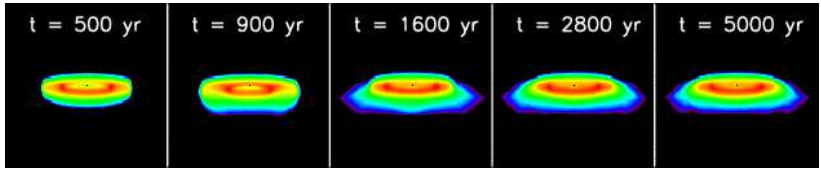}
\end{center}
\caption{Five snapshots of an initially unperturbed, near edge-on disk
  subjected to interstellar gas drag from a warm, low-density
  cloud. The model assumes a steady state develops such that each time
  a grain is lost from the initially unperturbed ring, a new
  replacement grain is generated within the initial ``birth ring.''
  The relative flow direction in this example is coplanar with the
  disk midplane, and the equilibrated disk is equivalent to the third
  column in the first row of Figure \ref{vr}.  The observed morphology
  does not change after $3 \times 10^3$ years, setting the timescale
  for which the system is assumed to have achieved steady state. \vspace*{2mm}}
\label{make_model_testequilibrate}
\end{figure}

To generate scattered-light images of the perturbed disk at some time
after the initial encounter with the cloud, we calculate the classical
elements from the resulting Runge-Lenz and angular momentum vectors
and spread each grain over 100 points in its orbit in proportion to
the time spent at each location.  We estimate the equilibration
timescale by monitoring the system morphology at successive times
after the initial encounter with the cloud.  Figure
\ref{make_model_testequilibrate} shows an example of the disk
morphology evolution for a highly inclined disk, $10^\circ$ from
edge-on.  The relative flow direction in this example is coplanar with
the disk midplane.  The observed morphology does not change after $3
\times 10^3$ years, setting the timescale for which the system is
assumed to have achieved steady state.

To create scattered light model images for comparison to HD 61005, we
incline the equilibrated disk by $80^\circ$ and $-80^\circ$.  We also
generate model images inclined by 0$^\circ$ (face-on), 45$^\circ$,
90$^\circ$ (edge-on), and $-45^\circ$ to illustrate the more general
consequences of neutral gas drag.  In our notation, positive
inclinations indicate that the lower half of the disk in the image is
closer to the observer than the upper half.  The pixel size in our
images is 2.6 AU, corresponding to the projected size of a NICMOS
pixel ($0.0759^{\prime\prime}$) at the distance of HD 61005 (34.5 pc).
The scattered light images also assume a Henyey-Greenstein phase
function with a scattering asymmetry typical of debris disks (g =
0.3; Meyer et al. 2007).  If the grains in HD 61005 are similar to
those in AU Mic, as suggested by the color and polarization results
(\S 3), a larger scattering asymmetry may be more appropriate (e.g., g
$\sim$ 0.7; Graham et al. 2007).  However, the adopted scattering
asymmetry has a relatively small effect on the qualitative morphology
for highly inclined outer disks like HD 61005, as the majority of
scattering angles present are near $90^\circ$.

The grid of models described above are shown in Figures \ref{vr},
\ref{vrvz}, and \ref{vz}. The image display scale is logarithmic, and each
model is $\approx$310 AU ($=9^{\prime\prime}$ for HD 61005) to a side.
The labels at top indicate the disk inclination, and the labels at
left indicate the azimuthal direction of the radial component of the
flow vector.  The bottom labels indicate the magnitude of the flow
vector parallel and perpendicular to the disk midplane.  The adopted
coordinate system is illustrated with respect to the face-on disks in
the left column of each Figure.

When a significant fraction of the relative flow vector is parallel to
the disk midplane, Figures \ref{vr} and \ref{vrvz} show that a variety
of morphologies can be produced, including bow structures
($i=0^\circ$), brightness asymmetries
($i=80^\circ,90^\circ,-80^\circ$), and warps ($i=90^\circ/
\theta=0^\circ$).  As shown analytically by \citet{Scherer00} and
discussed in \S 4.2.2, the neutral gas drag force acts to rotate the
grain pericenters into a direction perpendicular to the flow vector,
resulting in a build up of particles in that direction.  This effect
is clearly observed in the face-on cases ($i=0^\circ$) in Figures
\ref{vr} and \ref{vrvz}.

For a highly inclined disk and a small range of radial flow directions
($\theta \sim 0^\circ, 180^\circ$), the models in Figures \ref{vr} and
\ref{vrvz} show a swept morphology somewhat similar to HD 61005. This
effect is produced both by the build up of particles perpendicular to
the flow vector and the non-zero inclination of the disk.
Counterintuitively, the disk models with $v_r = 0$ and $v_z = |v|$
(Figure \ref{vz}) do not show this structure and instead produce a
largely symmetric distribution about the disk midplane.  As discussed
in \citet{Scherer00}, this is because the direction in which a given
grain's pericenter rotates to become perpendicular to the flow depends
on the initial pericenter direction. As a result, grains with an
initial direction to pericenter above the nominal midplane will
persist in having a direction to pericenter at or above the midplane
as long as they are bound to the star.

{}

\end{document}